\documentclass[reprint,notitlepage,
superscriptaddress,
nofootinbib,floatfix,aps,prd,
]{revtex4-2}
\pdfoutput=1
\usepackage{graphicx} % Include figure files
\usepackage{mathtools}
\usepackage[utf8]{inputenc} 
\usepackage{xcolor}
\usepackage{aas_macros}
\usepackage{url}
\usepackage[colorlinks]{hyperref}
\usepackage{calc}
\usepackage{amsmath}
\usepackage{amssymb}
\usepackage{amsthm}
\usepackage[bottom]{footmisc}
\usepackage[final]{changes}

\definechangesauthor[name=Giancarlo, color=red]{ }

\newcommand{\gcadd}[2][]{\added[id={ }, comment={#1}]{#2}}
\newcommand{\gcdel}[2][]{\deleted[id={ }, comment={#1}]{#2}}
\newcommand{\gcrep}[3][]{\replaced[id={ }, comment={#1}]{#2}{#3}}

\newcommand{\Hzero}{\ensuremath{{\cal H}_0}}
\newcommand{\Hone}{\ensuremath{{\cal H}_1}}
\newcommand{\pfa}{\ensuremath{P_{FA}}}
\newcommand{\pd}{\ensuremath{P_{D}}}
\newtheorem{theorem}{Theorem}[section]

\definecolor{dodgerblue}{HTML}{1E90FF}
\definecolor{ferrarired}{HTML}{ff2800}
\definecolor{olive}{HTML}{808000}
\definecolor{maroon}{HTML}{800000}
\newcommand\orcidlink[1]{\href{https://orcid.org/#1}{\includegraphics[scale=0.006]{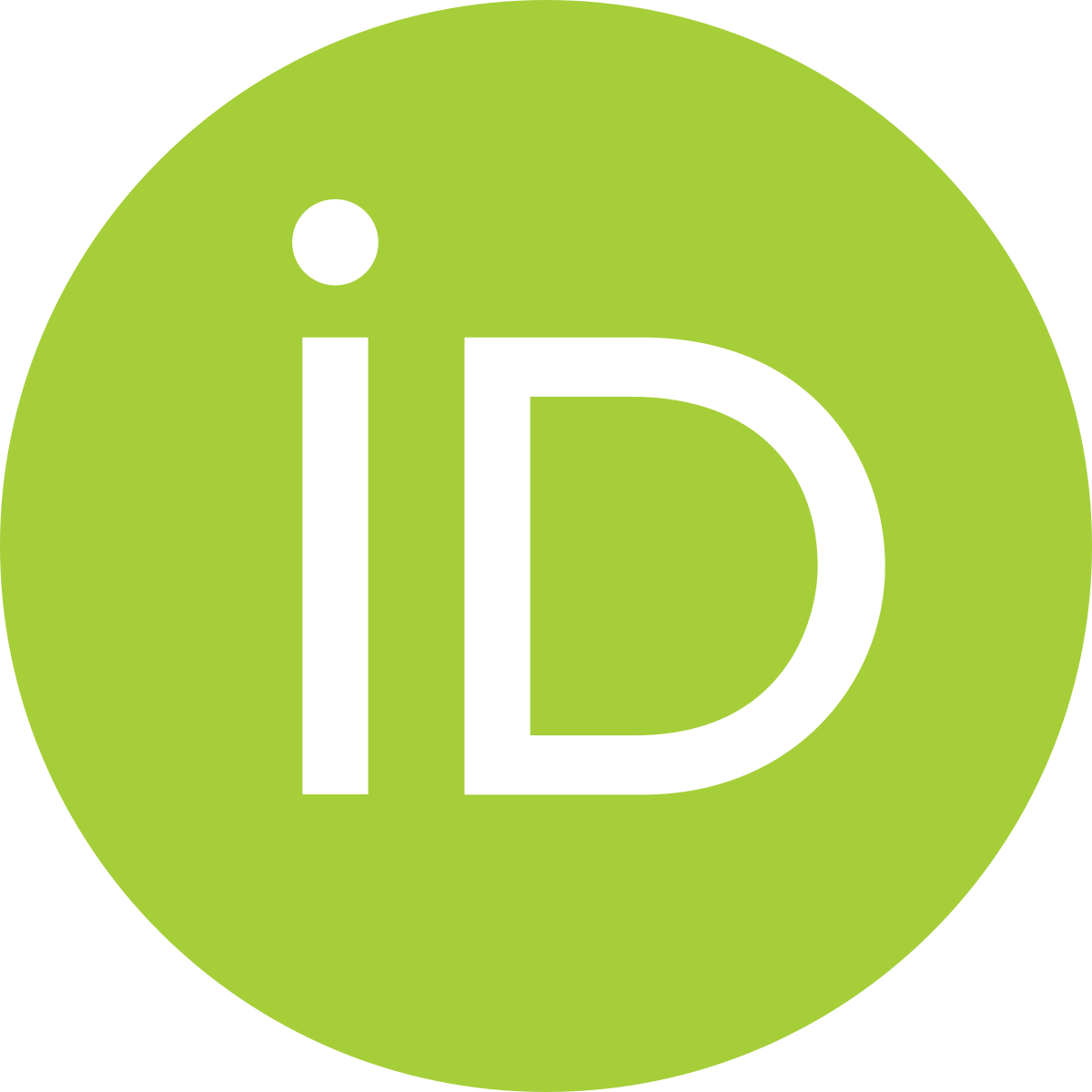}}}

\DeclareMathOperator{\erfc}{erfc}

\makeatletter
\newcommand{\distas}[1]{\mathbin{\overset{#1}{\kern\z@=}}}
\makeatother

\makeatletter
\def\namedlabel#1#2{\begingroup
	\def\@currentlabel{#2}%
	\label{#1}\endgroup}
\makeatother
\definecolor{LinkColor}{rgb}{0.75, 0, 0}
\definecolor{CiteColor}{rgb}{0, 0.5, 0.5}
\definecolor{UrlColor}{rgb}{0, 0, 0.75}
\hypersetup{linkcolor=LinkColor}
\hypersetup{citecolor=CiteColor}
\hypersetup{urlcolor=UrlColor}

\begin{document}

%\linenumbers
	\title{Improved detection statistics for non--Gaussian gravitational wave stochastic backgrounds}
	
	%Affiliations
	\newcommand{\infnpisa}{\affiliation{INFN Sez.~Pisa, Largo B. Pontecorvo 3, I-56127 Pisa, Italy}}
	\newcommand{\unipi}{\affiliation{Universit\`{a} di Pisa, Dipartimento di Fisica ``E. Fermi'', Largo B. Pontecorvo 3,  I-56127 Pisa, Italy}}
	\newcommand{\bham}{\affiliation{Institute for Gravitational Wave Astronomy \& School of Physics and Astronomy, University of Birmingham, Birmingham, B15 2TT, UK}}
	\newcommand{\milan}{\affiliation{Dipartimento di Fisica ``G. Occhialini'', Universit\'a degli Studi di Milano-Bicocca, Piazza della Scienza 3, 20126 Milano, Italy} \affiliation{INFN, Sezione di Milano-Bicocca, Piazza della Scienza 3, 20126 Milano, Italy}}
	
	\author{Matteo Ballelli~\orcidlink{0000-0003-1512-5423}}
	\infnpisa
	\unipi
	
	\author{Riccardo Buscicchio~\orcidlink{0000-0002-7387-6754}}
	\email{riccardo.buscicchio@unimib.it}
	\milan
	\bham
	
	\author{Barbara Patricelli~\orcidlink{0000-0001-6709-0969}}
	\infnpisa
	\unipi
	
	\author{Anirban Ain~\orcidlink{0000-0003-4534-4619}}
	\infnpisa
	\unipi
	
	\author{Giancarlo Cella~\orcidlink{0000-0002-0752-0338}}
	\infnpisa
	
	\date{\today}
	
	\begin{abstract}		
		In a recent paper we described a novel approach to the detection and parameter estimation of a non\textendash Gaussian stochastic background of gravitational waves.
		\gcdel{
		We devised an inference procedure that uses signal realizations and importance sampling to weight integrals appearing in relevant statistical quantities.
		In particular, we constructed the associated detection statistics: in order to provide robustness against stationary noise uncertainties we proposed a subtraction procedure to remove terms with non--zero expectation values in absence of signal.
		We characterized the detector statistics performances, and observed that for low to moderate non-Gaussianities it is outperformed by established Gaussian inference schemes.}
		In this work we propose \gcrep{an improved version of the detection procedure, preserving robustness against imperfect noise knowledge at no cost of detection performance:}{a more careful, robust subtraction procedure: while still using the importance sampling scheme, it does not introduce performance penalties.}
		\gcadd{in the previous approach, the solution proposed to ensure robustness reduced the performances of the detection statistics, which in some cases (namely, mild non--Gaussianity) could be outperformed by Gaussian ones established in literature}.
		\gcdel{We provide formal proof of its efficiency and, following closely the approach in our previous paper, we benchmark its performances on the same toy model: the proposed approach performs better than the Gaussian statistics everywhere in the model parameter space, therefore constituting a crucial addition to our framework.
		}
		\gcadd{We show, through a simple toy model, that} \gcrep{t}{T}he new detection statistic 
		\gcrep{performs better than the previous one (and than the Gaussian statistic) everywhere in the parameter space. 
		It approaches the optimal Neyman--Pearson statistics monotonically with increasing non--Gaussianity and/or number of detectors.}{performs better than the previous one (and better than the Gaussian statistic) everywhere in the parameter space . 
		It approaches the optimal Neyman--Pearson statistics when non--Gaussianity increases and/or for a large enough number of detectors.}
		\gcrep{In this study we discuss in detail its efficiency.}{A detailed discussion of new detection statistic's efficiency is given.}
		This is a \gcrep{second, important}{first} step towards the implementation of a nearly--optimal detection procedure for a realistic non--Gaussian stochastic background. We discuss the relevance of results obtained in the context of the toy model used, and their importance for understanding a more realistic scenario. 
	\end{abstract}
	
	\maketitle
	
	\section{Introduction}
	\label{sec:Introduction}
	
	The \gcrep{non--Gaussian}{astrophysical} stochastic gravitational wave (GW) background is an interesting and promising target for GW detectors.
	It originates from a superposition of many uncorrelated and unresolved events. 
	\gcrep{An interesting example is the}{ For a} background originating from compact binary coalescences, \gcrep{which is within reach of current generation detector network}{the opportunity to detect it with current detector network sensitivities is within reach}~\cite{2021PhRvD.104b2004A}, \gcadd{but several others can be observed in the future, both of astrophysical~\cite{TaniaRegimbau_2011} and cosmological origin~\cite{PhysRevLett.85.3761}.}
	
	The statistical properties of  \gcrep{a non--Gaussian}{an astrophysical} background are related to \gcadd{astrophysically relevant} quantities of great interest, such as the \gcdel{astrophysical} event rate and the population distribution of source parameters.
	In the presence of strong overlap between individual signals generated by the different events, the stochastic background \gcrep{is well modeled by}{can be described as} a Gaussian stochastic process, fully characterized by its second order statistic. 
	\gcrep{In absence of it}{When this is not true } \gcdel{(e.g. for a background signal arising from binaries of \gcrep{black holes}{neutron stars})}, the statistical distribution \gcadd{of the strain signals measured by the detectors} becomes non--Gaussian and contains larger information that can be in principle extracted. \gcadd{For example, accordingly with the rate estimates based on recent LVK observations \cite{2021arXiv211103634T}  the background generated by BBH coalescences is expected to be strongly non--Gaussian, with a product between the  length of the observable coalescence signal and the event rate  $O(10^{-3})$. The same parameter is $O(10)$ for BNS coalescences, and non--Gaussianity is expected to be less important.}

	Several detection strategies \gcadd{to efficiently detect a non--Gaussian background} have been proposed in literature: Drasco \& Flanagan \cite{Drasco:2002yd} devised a likelihood appropriate for \gcrep{the superposition}{superpositions} of burst--like signals; 
	therein an optimal detection statistics was derived \gcrep{assuming well separated, burst--like signals, coaligned and colocated detectors, and white noise;}{with some additional assumptions;}
	an approach applicable \gcrep{without these assumptions, but only with a negligible superposition probability for different events}{in more general and realistic scenarios} has been proposed by Thrane \cite{2013PhRvD..87d3009T};
	Smith \& Thrane \cite{Smith:2018PhRvX...8b1019S} \gcadd{further} proposed an optimal Bayesian parameter estimation method suitable for a background of astrophysical events in the low overlap regime. 
	
	In all the above proposals the basic building block is \gcrep{ the }{ a } \gcdel{probability distribution for the measured signal which
		can be understood as a probabilistic superposition of two
		terms, describing the presence and the absence of GWs
		in a given data segment.} probability distribution \gcadd{$p$ for a convenient statistics of the data $\rho$ that can be evaluated for each segment of a measured signal. This is written as a weighted sum}
	\begin{equation}
		p(\rho) = \xi p_1(\rho) + (1-\xi) p_0(\rho)
	\end{equation}
	\gcadd{
	where $\rho \sim p_1$ when the given segment contains an astrophysical event and $\rho \sim p_0$ otherwise. The weight $\xi$ is just the probability of having an event in the segment.}
	
	Other approaches, based on semiparametric models \cite{2014PhRvD..89l4009M} or higher order statistics \cite{2008ApJ...683L..95S,2009PhRvD..80d3003S} have also been explored. A comprehensive review can be found in \cite{Romano:2016dpx}.
	
	In a recent paper~\cite{PhysRevD.107.063027} we discussed a direct approach for detection and parameter estimation of a non-Gaussian stochastic background of GWs through a network of detectors.
	The core idea was to \gcrep{try to follow as much as possible an optimal approach. The target was to implement an optimal detection statistics (DS), and similarly a Bayesian approach to parameter estimation based on a realistic likelihood ${\cal L}$. To circumvent the lack of analytical closed--form expressions for the DS and for ${\cal L}$ we opted to \textit{numerically evaluate} them. This approach implies approximations of relevant integrals, that can be reduced at will with a large enough computational power.}{estimate the quantities
	of interest, i.e. the likelihood or the optimal detection statistic (DS), thus circumventing the lack of analytical closed-form expressions or the high computational cost required to evaluate them.}
	We proposed a Monte Carlo importance sampling procedure to achieve this objective. \gcadd{In this paper we focus on a further improvement of the DS, while not investigating parameter estimation, anymore. In Sec.~\ref{sec:summary} we summarize the our previous relevant findings, leading to the improvement presented in this paper.}
	
	\section{\gcadd{Summary of previous results}\label{sec:summary}}
	As a first step in~\cite{PhysRevD.107.063027} we \gcrep{studied the optimal Neyman--Pearson DS for }{ applied the proposed procedure to} a simple toy model, where the stochastic signal is a sequence of independent pulses with  amplitudes distributed as a Gaussian mixture. \gcdel{We evaluated the optimal DS analytically and characterized its detection performances.}  
	\gcrep{In this particular case}{In particular,} the DS can be evaluated in a closed form, as a nonlinear function of the data.
	
	A conceptual issue arised in this context: \gcrep{the optimal Neyman--Pearson}{the} DS, \gcadd{which is obtained under the assumption of known noise,} contains contributions proportional to single detector data autocorrelations (of arbitrary order). These contributions are \gcrep{ dominated by }{ sensitive to }  the noise~\footnote{We always assume a Gaussian noise.}, and in a realistic regime the noise amplitude \gcrep{cannot}{can't} be \gcrep{assumed known}{predicted} with a precision high enough to not spoil the \gcrep{DS}{ statistic } effectiveness. 
	
	\gcadd{The same shortcoming has been encountered in literature also in the Gaussian case, and we will clarify it with an example. Let us suppose we want to discriminate between the models}
	\gcadd{
	\begin{align}
		{\cal H}_0: \qquad s^{\cal A}_i &= \sigma n^{\cal A}_i \\
		{\cal H}_1: \qquad s^{\cal A}_i &= \sigma n^{\cal A}_i + A h_i
	\end{align}
	where $s^{\cal A}_i$ are measured values ($1<{\cal A}<N_D$ is an index labelling the detector and $i\leq 1\leq N$), $\sigma$ and $A$ are the noise and signal amplitudes. The stochastic variables $n_I$, $h$ are normally distributed and independent. Under the assumption of a known $\sigma^2$ the Neyman-Pearson test is easily found from
	\begin{equation}
		\frac{p(s_i^{\cal A}|\sigma^2,A^2 )}{p(s_i^{\cal A}|\sigma^2,0 )} > \lambda
	\end{equation}
	which is equivalent to
	\begin{equation}
		\label{eq:NPDS}
		Y = \frac{1}{N} \sum_{i=1}^N \left[ 2\sum_{\cal A>B} s^{\cal A}_i s^{\cal B}_i  
		+ \sum_{\cal A} \left(s^{\cal A}_i \right)^2 \right] > \lambda
	\end{equation}
	The mean of $Y$ is 
	\begin{equation}
		\mu_Y = \left[ N_D(N_D-1) A^2  + N_D (A^2+\sigma^2) \right].
	\end{equation}
	 We see that the first term, originating from cross-correlations, is noise independent while the second, originating from auto-correlations, depends both upon signal and noise. 
	 As a matter of fact for a realistic stochastic background $\sigma^2$ is uncertain, with an uncertainty larger than $A^2$. This means that we will not be able to detect confidently a stochastic background from the DS in~\eqref{eq:NPDS}: a given value of $Y$ could be explained both by a background contribution, or by a noise fluctuation. The problem disappears when the noise dominated diagonal components are removed, obtaining 
	\begin{equation}
	\label{eq:NPDSS}
	Y = \frac{1}{N} \sum_{i=1}^N \left[ 2\sum_{\cal A>B} s^{\cal A}_i s^{\cal B}_i  \right] > \lambda
	\end{equation}
	This is a very simplified version of the statistic commonly employed to detect a Gaussian stochastic background, see for example Eq.~(3.73) in~\cite{1999PhRvD..59j2001A}
	 which defines the optimal cross--correlation in the Gaussian case
	 \begin{equation}
	 	\label{eq:optAllenRomano}
	 	S \propto \int_{-\infty}^{\infty} \frac{\gamma(f) \Omega_{GW}(f) }{f^3 S_n^{\cal A}(f) S_n^{\cal B}(f)} s^{\cal A}(f)^* s^{\cal B}(f) df
	 \end{equation}
 }
	 \gcadd{A simplification comes from the assumption of co--located and co--oriented detectors which makes the overlap reduction function $\gamma(f)$ equal to one. Furthermore, both the stochastic signal and the noise are assumed here to have a white spectrum, so $\Omega_{GW}\propto f^3$, $S_n^{\cal A,B}$ are frequency independent and Eq.~(\ref{eq:optAllenRomano}) becomes equivalent to Eq.~(\ref{eq:NPDS}).
	}
\gcadd{
	It is interesting to note that if we use a generalized likelihood ratio test (GLRT) test like
	\begin{equation}
		\frac{p(s_i^{\cal A}|\hat{\sigma}^2_1,\hat{A}^2_1 )}{p(s_i^{\cal A}|\hat{\sigma}^2_0,0)} > \lambda
	\end{equation}
	where $\hat{\sigma}_i$ and $\hat{A}_i$ are the maximum likelihood estimates of $\sigma$ and $A$ under the ${\cal H}_1$ hypothesis we obtain, when $\sigma^2 \gg A^2$, 
	\begin{equation}
	Y = \sqrt{\frac{\hat{A}_1^2}{ \hat{\sigma}_0^2}} = \frac{\frac{1}{N} \sum_i \sum_{\cal A > B} s^{\cal A}_i s^{\cal B}_i}{\frac{1}{N} \sum_i \sum_{\cal A} (s^{\cal A}_i)^2} > \lambda
	\end{equation}
}
\gcadd{
For large enough $N$, relative fluctuations in the denominator become negligible, and we recover the DS in Eq.~\eqref{eq:NPDSS}.
}
   \gcdel{For a Gaussian stochastic background the optimal statistic  is a quadratic function of the data, and it is easy to remove contribution diagonal in detector indices just by removing products of data coming from the same detector.} 
	Under the assumption that noises across different detectors are uncorrelated, the modified statistic has by construction a zero expectation value in the hypothesis ${\cal H}_0$ (i.e. absence of signal), and a  detection procedure robust against noise mismodelling can be defined (see for example~\cite{1999PhRvD..59j2001A}, or~\cite{2022Galax..10...34R} and references therein). 
	
	Removing analogous contributions\footnote{Because of the analogy with the Gaussian case we will henceforth refer to these terms as \emph{diagonal}. The generalized definition of a \emph{diagonal} term is the following: a correlation between the data, of arbitrary order, that has a non zero average in the ${\cal H}_0$ hypothesis of no background.} from the nonlinear statistic\gcadd{,} appropriate for a non Gaussian background\gcadd{,} is tricky.
	In~\cite{PhysRevD.107.063027} we defined a procedure involving the subtraction of a set of auxiliary data constructed from the original ones. \gcadd{The basic idea was to construct a set of data streams (one for each detector) with the same autocorrelation but with zero cross--correlation among themselves and with the original set. This can easily be done, for example by introducing a time shift larger than the correlation length of the stochastic background. In this way autocorrelations can be independently estimated and subtracted.} 
		
	However our approach was suboptimal: 
	auxiliary data \gcadd{have by construction zero average estimates for cross correlation, but these estimates --which are subtracted from the ones of the original data--} carry additional fluctuations. 
	\gcrep{These are summed to the ones of the original set: as a consequence, we obtain more noisy DS and the detection performance is reduced.}{, hence they reduce the detection performances}
	\gcadd{
	In~\cite{PhysRevD.107.063027} this was evident in the regime where non--Gaussianity was not too high. It was quantified by the reduction of detection probability for a given false alarm, which in some cases was worse than the one of the Gaussian DS. We show again these results in this paper, comparing them with the improved ones (see Figures~\ref{fig:nd2},\ref{fig:nd3},\ref{fig:nd4},\ref{fig:nd5}). 	
	} 
	
	In this work we define a different, straightforward procedure to remove \emph{diagonal} contributions from the nonlinear statistics. 
	This procedure does not introduce additional fluctuations, and has \gcrep{improved performances compared to~\cite{PhysRevD.107.063027}, when measured in term of figures of merit related to detection probability, as will be discussed extensively in Section~\ref{sec:results}.}{and much better performances compared to~\cite{PhysRevD.107.063027}.}
	We apply the new procedure to the same toy model used previously, and we quantify the resulting improvement. 
	Finally, we show that the new procedure can be applied to a realistic background.
	
	Henceforth, we will discuss and compare a number of detections statistics, with the following naming convention:
	\begin{itemize}
		\item The ``Exact'' DS is 
		constructed via the Neyman--Pearson lemma \cite{1933RSPTA.231..289N}. 
		It is formally the optimal one, but the presence of \emph{diagonal} terms makes it unusable in a realistic setup, i.e. when uncertainty on the noise amplitude is larger than the target GW signal under the hypothesis ${\cal H}_1$.
		\item The ``Scrambled'' DS is obtained from the ``Exact'' statistic with the subtraction procedure defined and characterized in~\cite{PhysRevD.107.063027}\gcadd{, and shortly described in this Section.}
		\item The ``Improved'' DS is obtained from the ``Exact'' statistic with the new subtraction procedure described in \gcadd{Section~\ref{sec:Improved-detection}}.
		\item The ``Gaussian'' DS is the optimal Neyman--Pearson detector for a Gaussian stochastic background with a spectrum matching that of the toy model \gcadd{described in Section~\ref{sec:Improved-detection}, the same considered in}~\cite{PhysRevD.107.063027}. 
		It is suboptimal when applied to a non-Gaussian stochastic background, and is used as a fiducial reference \gcrep{for performance comparison.}{point.}
	\end{itemize}
	In this work we provide only minimal details on the ``Exact'', ``Scrambled'', and ``Gaussian'' DSs construction. 
	The interested reader will find relevant ones in~\cite{PhysRevD.107.063027} and references therein. 
	We will instead focus on the ``Improved'' statistics. 
	In Sec.~\ref{sec:Improved-detection} we construct it explicitly, providing a proof of its optimality with respect to a set of robustness requirements.
	In Sec.~\ref{subsec:perf-compar} we present performance comparisons between the above statistics, exploring their behaviour for varying signal duration, their amount of non-Gaussianity, and the number of detectors. 
	In Sec.~\ref{subsec:realistic}, we briefly discuss a realistic implementation of the procedure, highlighting its advantages.
	Finally, in Sec.~\ref{sec:Conclusions-and-perspectives} we draw conclusions and prospects for further developments of our approach.
	
	\section{Improved detection}
	\label{sec:Improved-detection}
	
	Our discussion will be mainly in the context of the toy model studied in~\cite{PhysRevD.107.063027}. 
	It allows for a quantitative comparison of detector performances. 
	In this model, the data measured by each detector are \gcadd{time series of the form}
	\begin{equation}
		\label{eq:signal_model}
		s_i^{\cal A} = n_i^{\cal A} + h_i \ ,
	\end{equation}
	\gcrep{where $i$ is a discrete time index, $1\leq i\leq N$, being $t_i=i \delta t$ the measurement time and $\delta t$ the sampling time. We suppose to have a network of $N_D$ detectors labeled by the index $\cal A$,  with $1\leq {\cal A}\leq N_D$.  The noise component of the data is $n_i^{\cal A}$, and  h}{where $n_i^{\cal A}$ is the $i$-th measured value of the ${\cal A}$ detector's noise, with $1\leq i \leq N$ and $1\leq {\cal A} \leq N_D$ Henceforth} we will assume that $n_i^{\cal A}$s are Gaussian, zero-average random numbers with covariances defined by
	\begin{equation}
		\label{eq:independent}
		\left< n_i^{\cal A} n_j^{\cal B}\right> = \sigma_{\cal A}^2 \delta_{ij} \delta^{\cal AB} \ .
	\end{equation}
	The signal $h_i$ is the same on each detector\gcrep{, so they can be thought as identical, aligned and co--located. I}{and i}t follows a Gaussian mixture model distribution
	\begin{align}
		h_i &\sim \sum_\alpha p_\alpha {\cal N}(\cdot \mid 0,\sigma_\alpha) \label{eq:hmodel}\\
		\sigma_h^2 &=\left<h_i^2\right> = \sum_\alpha p_\alpha \sigma_\alpha^2
		\ .
	\end{align}
	By hypothesis, different signal data points are therefore independent\gcadd{: the waveform associated to our events are very short, delta--like burst, without a resolved structure. Let us write the single event in the form
	\begin{equation}
	h_i(h,j) = h \delta_{ij}
	\end{equation}
	}
	\gcadd{We have two parameters: the amplitude $h$ and the index $j$ associated to the event time. As the model is time independent, all the values of $j$ have the same probability. If $p(h)$ is the probability distribution for the amplitude parameter, $h_i$ will be distributed as}
	\begin{equation}
		\label{eq:pop1}
		h_i \sim e^{-\Gamma \delta t} \left( \delta(\cdot) + \sum_{n=1}^\infty \frac{(\Gamma \delta t)^n}{n!} (\underbrace{p \star p \star \cdots \star p}_{ \text{$n$ terms}})(\cdot) \right)
	\end{equation}
	\gcadd{where $\Gamma$ is the event rate. This can be approximated by the toy model with two mixture components $\alpha \in \{+,-\}$ introduced in~\cite{PhysRevD.107.063027} if $p\sim {\cal N}(\cdot \mid 0,\sigma_+)$ and $\Gamma \delta t \ll 1$, in such a way that terms with $n>1$ can be neglected in Eq.~\eqref{eq:pop1}
	\begin{equation}
	\label{eq:pop1_approx}
	h_i \sim (1-\Gamma \delta t) \delta(\cdot) + \Gamma \delta t {\cal N}(\cdot \mid 0,\sigma_+) 
	\end{equation}
		  This correspond to $\sigma_-=0$, $p_+=\Gamma \delta t$ and $\sigma_+/\sigma_h=1/\sqrt{\Gamma \delta t}$. The parameter $\Gamma \delta t$ is the average number of events that contributes to a given $h_i$, a measure of overlap and Gaussianity. For our delta--like signals the approximation $\Gamma \delta t \ll 1$ is not a problem, because the sampling time $\delta t$ is a parameter unrelated to the physics and could be assumed to be small. The two main limits of the model are the lack of any time structure for the event waveform, which does not allow for relevant overlap structure, and the approximation of identical (co--aligned and co--located) detectors.   		
	}
	\gcadd{To accommodate for the general case our model should be interpreted as a superposition of an astrophysical contribution like the previous one, with $p\sim {\cal N}(\cdot \mid 0,\sigma_+)$, and of a Gaussian contribution with variance $\sigma_-^2$. In this case (again in the approximation $\Gamma \delta t \ll 1$) we obtain}
	\begin{equation}
	\label{eq:pop2_approx}
	h_i \sim (1-\Gamma \delta t) {\cal N}(\cdot \mid 0,\sigma_-) + \Gamma \delta t {\cal N}(\cdot \mid 0,\sqrt{\sigma^2+\sigma_-^2}) 
	\end{equation}
	\gcadd{with $\sigma_h=\sqrt{\sigma_-^2+\Gamma\delta t\sigma^2}$ and $\sigma_+= \sqrt{\sigma^2+\sigma_-^2}$.}
	\begin{table}
		\begin{center}
			\begin{tabular}{p{.15\columnwidth}|p{.5\columnwidth}|p{.35\columnwidth}} 
				Physical quantity & Definition & Model parameters \\
				\hline
				& & \\
				$\Gamma$ & Astrophysical background events~rate & $\frac{1}{\delta t} \frac{\sigma_h^2-\sigma_-^2}{\sigma_+^2-\sigma_-^2}=\frac{p_+}{\delta t}$ \\
				$S_{h,Astro}$ & Astrophysical background power~spectrum & $(\sigma_+^2-\sigma_-^2)\delta t$ \\
				$S_{h,Gauss}$ & Gaussian background power~spectrum & $\sigma_-^2 \delta t$ \\		
				$T$ & Observing time & $N \delta t$		
			\end{tabular}			
		\caption{Relation between physical quantities and model parameters.}
		\label{tab:table1}
		\end{center}
	\end{table}
	The Neyman-Pearson optimal DS \gcadd{for the two component mixture model}, $\hat{Y}$,\gcdel{fot the model with two mixture components $\alpha \in {+,\times}$ introduced in [9]}  reads 
	\begin{equation}
		\label{eq:Y}
		\hat{Y} = \sum_{i} \hat{y}(w_i) \ ,
	\end{equation}
	where
	\begin{align}
		\hat{y}(w) & = \log \left[
		\sum_\alpha \frac{p_\alpha \sigma}{\sqrt{\sigma^2+\sigma_\alpha^2}}
		\exp \left(
		\frac{1}{2} \frac{\sigma_\alpha^2 w}{\sigma^2+\sigma_\alpha^2}
		\right)
		\right]  \ , \\
		w_i & = \sum_{\cal A B} u_i^{\cal A} u_i^{\cal B}   \ , \\
		u_i^{\cal A} &= \frac{\sigma s_i^{\cal A}}{\sigma_{\cal A}^2}  \ , \\
		\frac{1}{\sigma^2} & = \sum_{\cal A} \frac{1}{\sigma^2_{\cal A}} \ .
	\end{align}
	We can expand the DS as
	\begin{align}
		\hat{y}(w) 
		& = \sum_{n=0}^{\infty} \frac{1}{n!} \hat{y}^{(n)}(0) \left(\sum_{{\cal A} } u^{{\cal A}} \right)^{2n} \\ 
		&= \sum_{n=0}^{\infty} \frac{1}{n!} \hat{y}^{(n)}(0) Z(n,N_D)  \ , \label{eq:exp}
	\end{align}
	with
	\begin{align}
		\label{eq:ZnNd}
		Z(n,N_D) & =\!\!\! \! \!\!\!\!\!\!\sum_{k_1+\cdots+k_{N_{D}}=2n}
		\!\binom{2n}{k_1,\cdots,k_{N_D}}\!\! \prod_{{\cal A}=1}^{N_D} (u^{\cal A})^{k_{\cal A}} \ ,
	\end{align}
	where $k_{\cal A}$ are non-negative integers.
	Following the argument provided in Sec.~\ref{sec:summary} we want to remove from Eq.~\eqref{eq:exp} all terms with non-zero expectation value under the hypothesis ${\cal H}_0$ (i.e., in absence of GW signal). 
	Under such hypothesis, the $u^{\cal A}$s are independent normal variables and by virtue of Isserlis' theorem the terms to be cancelled are those in the large sum of~\eqref{eq:ZnNd} with all even $k_{\cal A}$s.
	
	This can be written explicitly, up to to an irrelevant multiplicative constant, as
	\begin{align}
		\label{eq:subtraction}
		\hat{y}_s &= \hat{y}\left[\left(\sum_{\cal A} u^{\cal A}\right)^{2}\right]  \nonumber\\
		& -
		\frac{1}{2^{N_D}}\!\!\! \sum_{\varepsilon_1 = -1,1} \cdots \!\!\!\! \sum_{\varepsilon_{N_D}= -1,1} \hat{y}\left[ \left( \sum_{{\cal A}} \varepsilon_{\cal A} u^{\cal A}\right)^{2}\right]\ .
	\end{align}
	A formal proof that this is the desired DS follows. 
	\begin{theorem}
		For a given statistics ${\cal S}(u^1,\cdots,u^{N_D})$, where $u^{\cal A}=h^{\cal A}+n^{\cal A}$, if $n^{\cal A}$ are statistically independent stochastic variables (i.e., the noise in our context) with non-zero even momenta and zero odd momenta, we consider the formal Taylor expansion in powers of $u^{\cal A}$. 
		A modified statistics where under $\mathcal{H}_0$ terms with non-zero expectation values are canceled while preserving others is
		\begin{align}
			\label{eq:subtracted}
			{\cal S}_s(u^1,\cdots,u^{N_D}) &= {\cal S}\left(u^1,\cdots,u^{N_D}\right) \,\,+ \nonumber\\
			-\frac{1}{N_D} \! \sum_{\varepsilon_1 = -1,1} \cdots \!\sum_{\varepsilon_{N_D}=-1,1} &{\cal S}\left(\varepsilon_1 u^1,\cdots,\varepsilon_{N_D} u^{N_D}\right) \ .
		\end{align} 
	\end{theorem}
	\begin{proof}
		Let us consider a generic term of the Taylor expansion, which reads
		\begin{equation}
			\left( u^1 \right)^{k_1} \cdots \left( u^{N_D} \right)^{k_{N_D}}  \ ,
		\end{equation}
		where $k_i$ are positive indices. If $h^{\cal A}=0$\gcadd{,} its expectation value is different from zero iff all the $k_i$s are even. In fact, 
		\begin{equation}
			\left< \left( n^1 \right)^{k_1}\! \cdots \left( n^{N_D} \right)^{k_{N_D}}\!\! \right> \!=\! \left< \left( n^1 \right)^{k_1} \right> \cdots \left< \left( n^{N_D} \right)^{k_{N_D}}\! \right>  \ . 
		\end{equation}
		as the $n^{\cal A}$ are statistically independent. We observe that
		\begin{align}
			\frac{1}{N_D}  \sum_{\underline{\varepsilon}} \left( \varepsilon_1 n^1 \right)^{k_1}\! \cdots \left( \varepsilon^{N_D}  n^{N_D} \right)^{k_{N_D}}   = \nonumber\\
			\left\{\frac{1}{N_D}  \sum_{\underline{\varepsilon}} \varepsilon_1^{k_1}\cdots\varepsilon_{N_D}^{k_{N_D}} \right\}
			\left( n^1 \right)^{k_1} \cdots \left( n^{N_D} \right)^{k_{N_D}}  \ ,
		\end{align} 
		and the expression between curly braces is equal to one iff all the $k_i$ are even, and equal to zero otherwise. 
		Therefore the statistics in Eq.~\eqref{eq:subtracted} cancels the fully even terms, while preserving all the others.
	\end{proof}
	It is worth noting that the Gaussian noise assumption is not mandatory: 
	independence and zero odd momenta are required, only.
	Gaussian, zero average noise is a particular case. 
	We focus on two particular examples, $N_D=2,3$. 
	For $N_D=2$ we get
	\begin{equation}
		\hat{y}_s  = \frac{1}{2}\hat{y}\left[(u^1+u^2)^2\right]-\frac{1}{2}\hat{y}\left[(u^1-u^2)^2\right]\ .
	\end{equation}
	The first two non--zero orders in a power expansion are given by
	\begin{align}
		\hat{y}_s &= 2 \hat{y}_s^\prime(0) u^1 u^2  + \nonumber \\
		&+ 2 \hat{y}_s^{\prime\prime}(0)\left[ (u^1)^3 u^2 + u^1 (u^2)^3 \right] + O\left((u)^6\right)\ ,
	\end{align}
	and we see that for a linear function $\hat{y}$  the usual Gaussian optimal DS is recovered, while higher-order corrections are proportional to higher order correlation of the measured data. 
	As desired, \emph{diagonal} terms such as $(u^1)^2$, $(u^1)^2 (u^2)^2$ etc. are canceled under the null-hypothesis. 
	
	We also note here that under the competing hypothesis ($h\neq0$), $O(u^4)$ terms proportional to $h^2 n^2$ and to $h^4$ are preserved. 
	The first provide additional information on the GW spectra, boosted by the noises' spectra frequently assumed to be much larger. 
	The other contains information about non--Gaussianity. 
	Similar contributions will appear at higher orders.
	
	In the same way for $N_D=3$ we have
	\begin{align}
		\hat{y}_s  &= 
		\frac{3}{4}\hat{y}\left[(u^1+u^2+u^3)^2\right] -\frac{1}{4}\hat{y}\left[(-u^1+u^2+u^3)^2\right]+\nonumber\\
		&-\frac{1}{4}\hat{y}\left[(u^1-u^2+u^3)^2\right]-\frac{1}{4}\hat{y}\left[(u^1+u^2-u^3)^2\right]\!\!\!\ ,
	\end{align}
	and the first two non-zero orders are
	\begin{align}
		\hat{y}_s 
		& = 2 \hat{y}_s^\prime(0) \left[u^1 u^2 + u^2 u^3 + u^3 u^1\right]
		+\nonumber \\
		&+ 2 \hat{y}_s^{\prime\prime}(0) 
		\left[3 (u^1)^2 u^2 u^3 + 3 u^1 (u^2)^2 u^3 + 3 u^1 u^2 (u^3)^2 \right. 
		+& \nonumber \\
		& + \left. (u^1)^3 (u^2+u^3)\!+\! (u^2)^3 (u^1+u^3)\! +\! (u^3)^3 (u^1+u^2)  \right] \!
		+\nonumber \\
		& 
		+ O\left((u)^6\right)\ .
	\end{align}
	We get again the Gaussian optimal DS at the lowest order, which is just a sum over all pairings of the weighted cross correlations
	\begin{equation}
		\label{eq:gauss3}
		u^1 u^2 + u^2 u^3 + u^3 u^1 = \sigma^2 \left( \frac{s^1 s^2}{\sigma_1^2 \sigma_2^2} + \frac{s^2 s^3}{\sigma_2^2 \sigma_3^2} + \frac{s^3 s^1}{\sigma_3^2 \sigma_1^2}\right)\!\!\!\ ,
	\end{equation}
	and no \emph{diagonal} higher order terms.
	
	\section{Results}
	\label{sec:results}
	
	\gcadd{In order to contextualize the performances of the improved DS, we first discuss the connection between the toy model used for the tests and a realistic stochastic background. Let us remind that the former can be described as a superposition of statistically independent events.}
	\gcadd{Each event will produce a contribution to the strain }
	\begin{equation}
		h_{\mu\nu}(\boldsymbol{x},t) = u_{\mu\nu}(\boldsymbol{x},t-\tau_I,\boldsymbol{\lambda}_I)  
	\end{equation}
	\gcadd{where $\tau_I$ is the event time and $\boldsymbol{\lambda}_I=(\lambda_I^1,\cdots,\lambda_I^{N_p})$ an appropriate set of parameters describing the event. }
	
	\gcadd{These can be intrinsic (e.g., the chirp mass of a coalescing binary) or extrinsic (e.g., the source luminosity distance or its position in the sky). It is convenient for our modelling to isolate the arrival time and not describe it on the same footing of other parameters.}
	
	\gcadd{An useful formalism for the description of a sequence of independent events with a fixed rate is based on the introduction of a function $Q(\boldsymbol{\lambda})$ \cite{Kampen1992}.} 
		
	\gcadd{We will not give the explicit derivations (see \cite{Kampen1992} for details): the interesting final result is that the statistical cumulants of strain field can be written as}
	\begin{eqnarray}
		\label{eq:cumulants}
		\left<\left<h_{\mu_1\nu_1}(\boldsymbol{x}_1,t_1) \cdots h_{\mu_n\nu_n}(\boldsymbol{x}_n,t_n)\right>\right>  &= \\
		\int d\tau \int d\boldsymbol{\lambda} Q(\boldsymbol{\lambda}) &\times \\
		 \times u_{\mu_1\nu_1}(\boldsymbol{x}_1,t_1-\tau,\boldsymbol{\lambda}) \cdots u_{\mu_n\nu_n}(\boldsymbol{x}_n,t_n-\tau,\boldsymbol{\lambda})
	\end{eqnarray}
	\gcadd{where we assumed that the average value $\left<h_{\mu\nu}(\boldsymbol{x},t) \right>$ is zero.
	The quantity $Q(\boldsymbol{\lambda})d\boldsymbol{\lambda}$ can be interpreted as the rate of events in a given infinitesimal volume of the parameter space. This is more evident if we write}
	\begin{equation}
		Q(\boldsymbol{\lambda}) = \Gamma P(\boldsymbol{\lambda}) 
	\end{equation}
	\gcadd{where $\Gamma$ is the event rate and $P(\boldsymbol{\lambda})$ the probability density of $\lambda$ over its parameter space. We use this notation to also rewrite \eqref{eq:cumulants} as}
	\begin{multline}
		\label{eq:cumulants_time}
		\left<\left<h_{\mu_1\nu_1}(\boldsymbol{x}_1,t_1) \cdots h_{\mu_n\nu_n}(\boldsymbol{x}_n,t_n)\right>\right>  = \\ 
		\Gamma \int d\tau \overline{u_{\mu_1\nu_1}(\boldsymbol{x}_1,t_1-\tau,\boldsymbol{\lambda}) \dots u_{\mu_n\nu_n}(\boldsymbol{x}_n,t_n-\tau,\boldsymbol{\lambda})}
	\end{multline}
	\gcadd{or, in the frequency domain,}
	\begin{multline}
		\label{eq:cumulants_freq}
		\left<\left<\tilde h_{\mu_1\nu_1}(\boldsymbol{x}_1,\omega_1) \dots \tilde h_{\mu_n\nu_n}(\boldsymbol{x}_n,\omega_n)\right>\right>  = \\
		2 \pi \Gamma \delta(\omega_1+\dots+\omega_n) 
		 \overline{\tilde u_{\mu_1\nu_1}(\boldsymbol{x}_1,\omega_1,\boldsymbol{\lambda}) \dots \tilde u_{\mu_n\nu_n}(\boldsymbol{x}_n,\omega_n,\boldsymbol{\lambda})}
	\end{multline}
	\gcadd{Here $\overline{X}$ is the average over the parameters $\boldsymbol{\lambda}$ of $X$.}
	
	\gcadd{Note that, from an observational point of view, we are interested in cumulants of the strain measured by the detectors, namely}
	\begin{equation}
		\label{eq:meas_strain}
		h^{\cal A}(t) = \int D^{\cal A}_{ij}(t,t^\prime) h_{ij}(t^\prime,\boldsymbol{x}^{A}(t^\prime)) dt^\prime
	\end{equation}
	\gcadd{Here $D^{A}_{ij}$ is a generalized detector tensor of the $A$--th interferometer located in $\boldsymbol{x}^A$, where we included the time dependency connected to the detector´s movement and a whitening transformation that can be chosen in such a way to normalize the signal to a white noise spectrum. 
	Moments of measured strains defined in Eq.~\eqref{eq:meas_strain} can be easily recovered from \eqref{eq:cumulants}. }
	
	\gcadd{All the information of astrophysical interest are encoded in $Q(\boldsymbol{\lambda})$. $Q$ functions are additive: if a given stochastic background is the sum of several contributions, each described by $Q^{(a)}(\boldsymbol{\lambda})$, we have}
	\begin{equation}
		Q(\boldsymbol{\lambda}) = \sum_a Q^{(a)}(\boldsymbol{\lambda})
	\end{equation}
	
	\gcadd{This mathematical description let us highlight an important fact: let us suppose that the rate is increased $Q(\boldsymbol{\lambda})\rightarrow \eta Q(\boldsymbol{\lambda})$, and at the same time individual events amplitudes are scaled accordingly with $u_{\mu\nu}\rightarrow \eta^{-1/2} u_{\mu\nu}$: the second order cumulant $\left<\left<h_{\mu_1\nu_1}(\boldsymbol{x}_1,t_1) h_{\mu_2\nu_2}(\boldsymbol{x}_2,t_2)\right>\right>$ does not change, while higher order ones are rescaled as}
	\begin{multline}
		\left<\left<h_{\mu_1\nu_1}(\boldsymbol{x}_1,t_1) \cdots h_{\mu_n\nu_n}(\boldsymbol{x}_n,t_n)\right>\right> \\ \rightarrow \eta^{(2-n)/2}\left<\left<h_{\mu_1\nu_1}(\boldsymbol{x}_1,t_1) \cdots h_{\mu_n\nu_n}(\boldsymbol{x}_n,t_n)\right>\right>
	\end{multline}
	\gcadd{and become less and less important in the $\eta \rightarrow \infty$ limit. This offers a straightforward interpretation: when the background is generated by a very large number of weak events, it becomes a Gaussian stochastic field: only its second order cumulant, which is related to spectral characteristics, is relevant.}
	
	\gcadd{Let us now connect the formalism above with the proposed toy model. Considering $\sigma_-=0$ (Eq.~\eqref{eq:pop1_approx}) we have
	\begin{equation}
	\tilde h(\omega) = h \delta t
	\end{equation}
	and we can use directly Eq.~\eqref{eq:cumulants_freq} to get the even cumulants
	\begin{multline}
		\left<\left<\tilde h^{{\cal A}_1}(\omega_1) \cdots \tilde h^{{\cal A}_n}(\omega_n) 
		\right>\right> = 2\pi \Gamma \delta(\omega_1+\cdots+\omega_n) \overline{h^n} \\
		= 2 \pi \Gamma  (n-1)!! p_+ \sigma_+^n \delta t^n \delta(\omega_1+\cdots+\omega_n) 
	\end{multline}
	while the odd ones are zero. The general case in Eq.~\eqref{eq:pop2_approx} can be obtained by substituting in this result $\sigma_+^2 \rightarrow \sigma_+^2-\sigma_-^2$ and a by adding the contribution of Gaussian background with variance $\sigma_-^2$. This contributes only to the $n=2$ cumulant and we get
	\begin{multline}
		\left<\left<\tilde h^{{\cal A}_1}(\omega_1) \cdots \tilde h^{{\cal A}_n}(\omega_n) 
		\right>\right> \\
		= 2 \pi \Gamma  (n-1)!! \left( p_+ (\sigma_+^2-\sigma_-^2)^{n/2}+ \sigma_-^2  \delta_{n,2} \right) \\
		\delta t^n \delta(\omega_1+\cdots+\omega_n) 
	\end{multline}
    In particular the power spectrum is flat and given by
    \begin{equation}
    	S_h(f) = \sigma_h^2 \delta t
    \end{equation}
    which correspond to an adimensional energy density for logarithmic interval of frequency
    \begin{equation}
    	\Omega_{GW}(f) = \frac{4\pi^2}{3 H_0^2} f^3 \sigma_h^2 \delta t
    \end{equation}
	It is important to note that the distribution of signal amplitudes for each event depends on the specific details of $Q(\boldsymbol{\lambda})$: e.g., it is sensitive to the redshift $z$ distribution of the sources. The signal power spectrum depends only on $\sigma_h$, while higher order momenta carry information, and therefore can distinguish additional signal properties.}
	 
	\gcadd{The noise power spectrum can be written similarly as $S_n^{\cal A}(f) = \sigma^2_{\cal A} \delta t$. Finally, we write the signal--to--noise ratio for a pair of detectors as a function of the toy model parameters
	\begin{equation}
		\text{SNR}^2 = \frac{8}{25} T \Delta f \frac{\sigma_h^4}{\sigma_1^2 \sigma_2^2} 
	\end{equation}
	where $T$ is the observing time and $\Delta f$ the sensitivity bandwidth of the detectors. We stress that this is an appropriate figure of merit for the performances of the Gaussian analysis.}

	In order to compare DSs' performances, we follow closely the procedure in~\cite{PhysRevD.107.063027}. 
	In addition, we observe that momenta of the DS can be evaluated as an explicit integral, therefore we avoid a direct simulation of the data under $\mathcal{H}_1$. 
	For the ``Exact'' DS we write
	\begin{align}
		\label{eq:mean_Y_improved}
		\left< \hat{y}^K \right> & \!= 
		\int \cdots \int \prod_{\cal A} d s^{\cal A} 
		\log
		\left\{
		\sum_{\alpha=+,-} 
		\frac{p_\alpha\sigma}{\sqrt{\sigma_\alpha^2+\sigma^2}} \times \right. \nonumber\\
		&\!\!\times\left.
		\exp\left[ 
		\frac{\sigma_\alpha^2}{2(\sigma_\alpha^2+\sigma^2)} 
		\left(  
		\sigma \sum_A \frac{s^{\cal A}}{\sigma_{\cal A}^2} 
		\right) 
		\right] 
		\right\}^K \times \nonumber \\
		&\!\!\!\!\! \times \!\left[  
		\sum_{\alpha=+,-} 
		\frac{p_{\alpha}e^{-\frac{1}{2} [C_{\alpha}^{-1}]^{{\cal AB}} s^{\cal A} s^{\cal B}}}{\sqrt{(2\pi)^{N_D} \det C_\alpha}} \right]\ ,
	\end{align}
	where $C_{\pm}$ are matrices in detector space with entries defined by 
	\begin{align}
		C_\pm^{{\cal AB}}  = \sigma_{\cal A}^2 \delta_{{\cal AB}} \qquad & \text{under ${\cal H}_0$} \\
		C_\pm^{{\cal AB}}  = \sigma_{\cal A}^2 \delta_{{\cal AB}} + \sigma_\pm^2 \qquad & \text{under ${\cal H}_1$}\ .
	\end{align}
	and they describe the two multivariate Gaussian distributions entering the competing hypothesis.
	Similar expressions can be obtained for the other DSs listed in Sec.~\ref{sec:Introduction} (detailed derivations are provided in~\cite{PhysRevD.107.063027}).
	
	\subsection{Performance comparison}
	\label{subsec:perf-compar}
	
	When a series of $N_d$ data is collected from each detector, the DS is the sum of Eq.~\eqref{eq:Y} and for large enough $N_d$, $\hat{Y}$ is approximately Gaussian distributed. 
	Using this property and the connected momenta being proportional to $N_d$, i.e.
	\begin{align}
		\left. \mu \right|_{N_{d}=N} & = \left.N \mu \right|_{N_{d}=1} \ ,\\ 
		\left. \sigma^2 \right|_{N_{d}=N} & = \left. N \sigma^2 \right|_{N_{d}=1} \ ,
	\end{align}
	we write a relation between false alarm probability, detection probability, and $N$ in the form
	\begin{align}
		\label{eq:scaledpower}
		P_D 
		& = \frac{1}{2} \erfc 
		\left( 
		\left. 
		\frac{\sigma_{{\cal H}_0}}{\sigma_{{\cal H}_1}} \right|_{N_{d}=1}
		\erfc^{-1}(2P_{FA}) + \right. \nonumber\\
		& 
		\left. - \sqrt{\frac{N}{2}} 
		\left. 
		\frac{\mu_{{\cal H}_1} -  \mu_{{\cal H}_0}}{\sigma_{{\cal H}_1}} 
		\right|_{N_{d}=1}  
		\right)\ .
	\end{align}
	\gcadd{We note that this relation is valid for all the DS considered, for a large enough number of data $N_d$ (or, equivalently, for a large enough data taking time $T=\delta t N_d$). It follows that a naive definition of a signal--to--noise ratio 
	\begin{equation}
		\mathrm{SNR} = \frac{\mu_{{\cal H}_1} -  \mu_{{\cal H}_0}}{\sigma_{{\cal H}_1}}
	\end{equation}	
	can always be used. However, it is also clear that this SNR does not describes completely the performances of a given DS, as the ratio $\frac{\sigma_{{\cal H}_0}}{\sigma_{{\cal H}_1}}$ is also relevant. 
	For this reason, we argue that the best and unambiguous figure of merit for a comparison between the Gaussian and the non--Gaussian case is the value of the detection probability.
	}

	\begin{figure*}
		\centering
		\includegraphics[width=1.84\columnwidth]{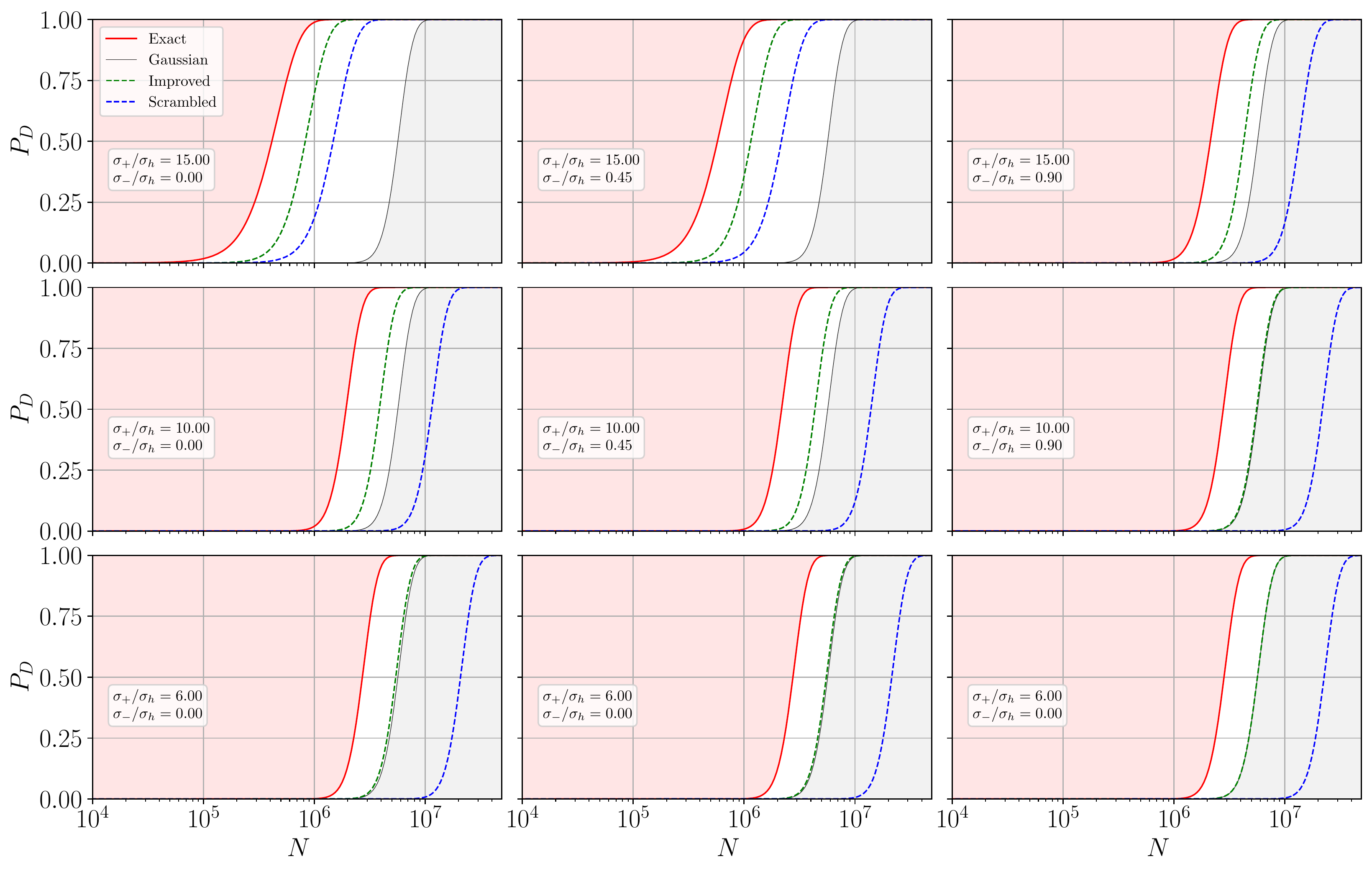}
		\caption{Comparison between detections statistics: 
			each panel contains the probability of detection $P_D$ as a function of the number of datapoints $N$. We show the ``Exact'', ``Improved'', ``Scrambled'', and ``Gaussian'' statistic as solid red, dashed green, dashed blue, and solid black lines, respectively. Several diffent signal models are considered, parameterized by $\sigma_+/\sigma_h$ and $\sigma_-/\sigma_h$, of increasing non-Gaussianity, going from bottom right to top left panel.
			We fix $N_D=2$ and $P_{FA}=10^{-15}$. 
			Shaded red areas delimit regions with beyond-optimal performances, right-bounded by the ``Exact'' DS. Shaded grey area delimit statistics with performances worse than the ``Gaussian'' one. 
			We observe uniform improvement over the whole parameter space of the ``Improved'' statistics as compared to the Gaussian one. 
			By contrast, the scrambling procedure introduced in~\cite{PhysRevD.107.063027} outperforms the Gaussian one for strong non-Gaussianities, only. 
			In Fig.~\ref{fig:nd3},~\ref{fig:nd4}, and~\ref{fig:nd5} similar results for increasing number of detectors are shown, with the ``Improved'' statistics approaching the ``Exact'' one.
		}
		\label{fig:nd2}
	\end{figure*}
	
	\begin{figure*}
		\centering
		\includegraphics[width=1.84\columnwidth]{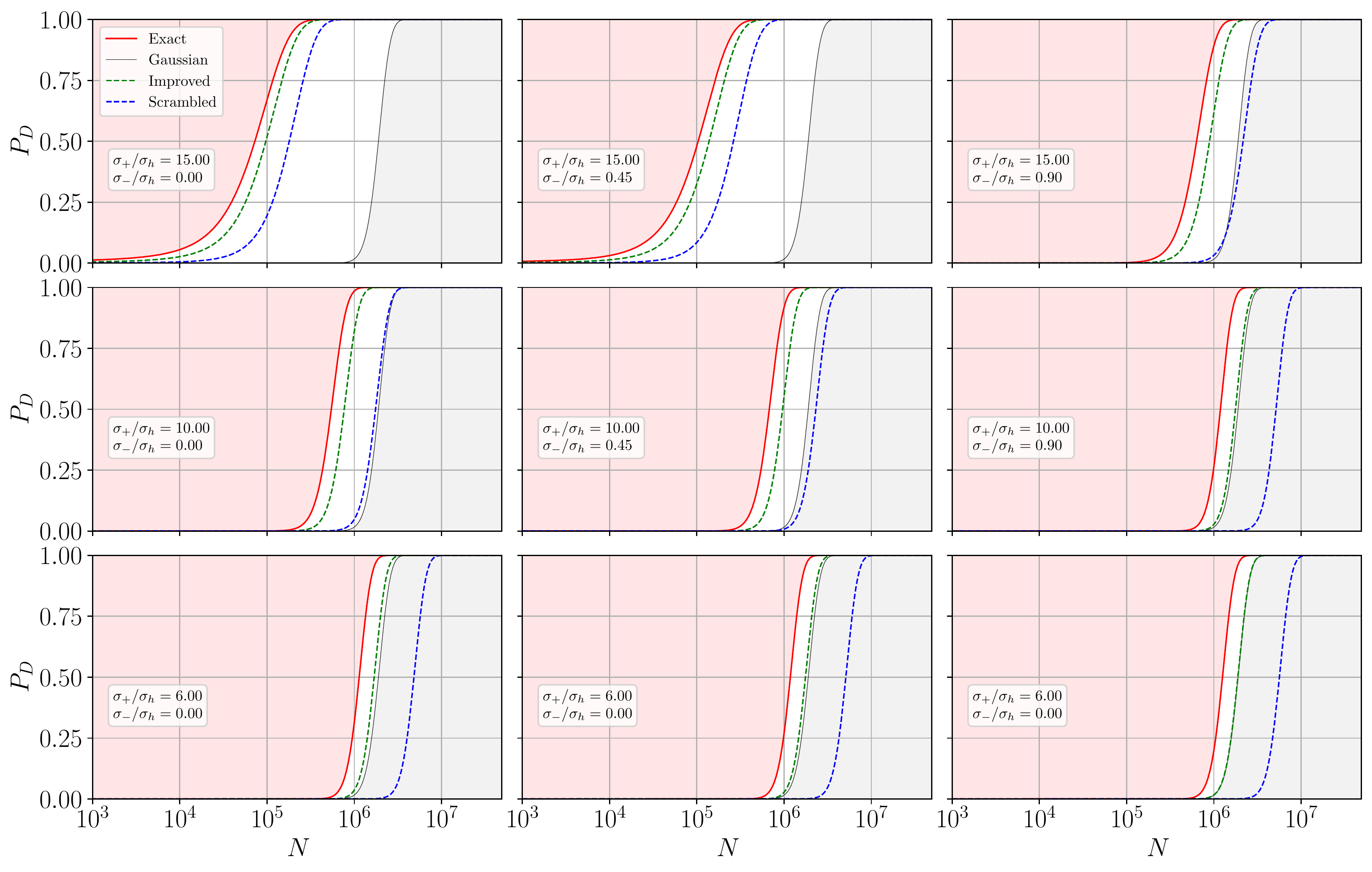}
		\caption{Comparison between detections statistics: each panel contains the probability of detection $P_D$ as a function of the number of datapoints $N$.
		We show the ``Exact'', ``Improved'', ``Scrambled'', and ``Gaussian'' statistic as solid red, dashed green, dashed blue, and solid black lines, respectively.
		Several diffent signal models are considered, parameterized by $\sigma_+/\sigma_h$ and $\sigma_-/\sigma_h$, of increasing non-Gaussianity, going from bottom right to top left panel.
		We fix $N_D=3$ and $P_{FA}=10^{-15}$. 
		Shaded red areas delimit regions with beyond-optimal performances, right-bounded by the ``Exact'' detections statistic. 
		Shaded grey areas delimit statistics with performances worse than the ``Gaussian'' one. 
		We observe uniform improvement over the whole parameter space of the ``Improved'' statistics as compared to the ``Gaussian'' one. 
		By contrast, the scrambling procedure introduced in~\cite{PhysRevD.107.063027} outperforms the Gaussian one for strong non-Gaussianities, only.\label{fig:nd3}
		}
	\end{figure*}
	
\begin{figure*}
		\centering
		\includegraphics[width=1.84\columnwidth]{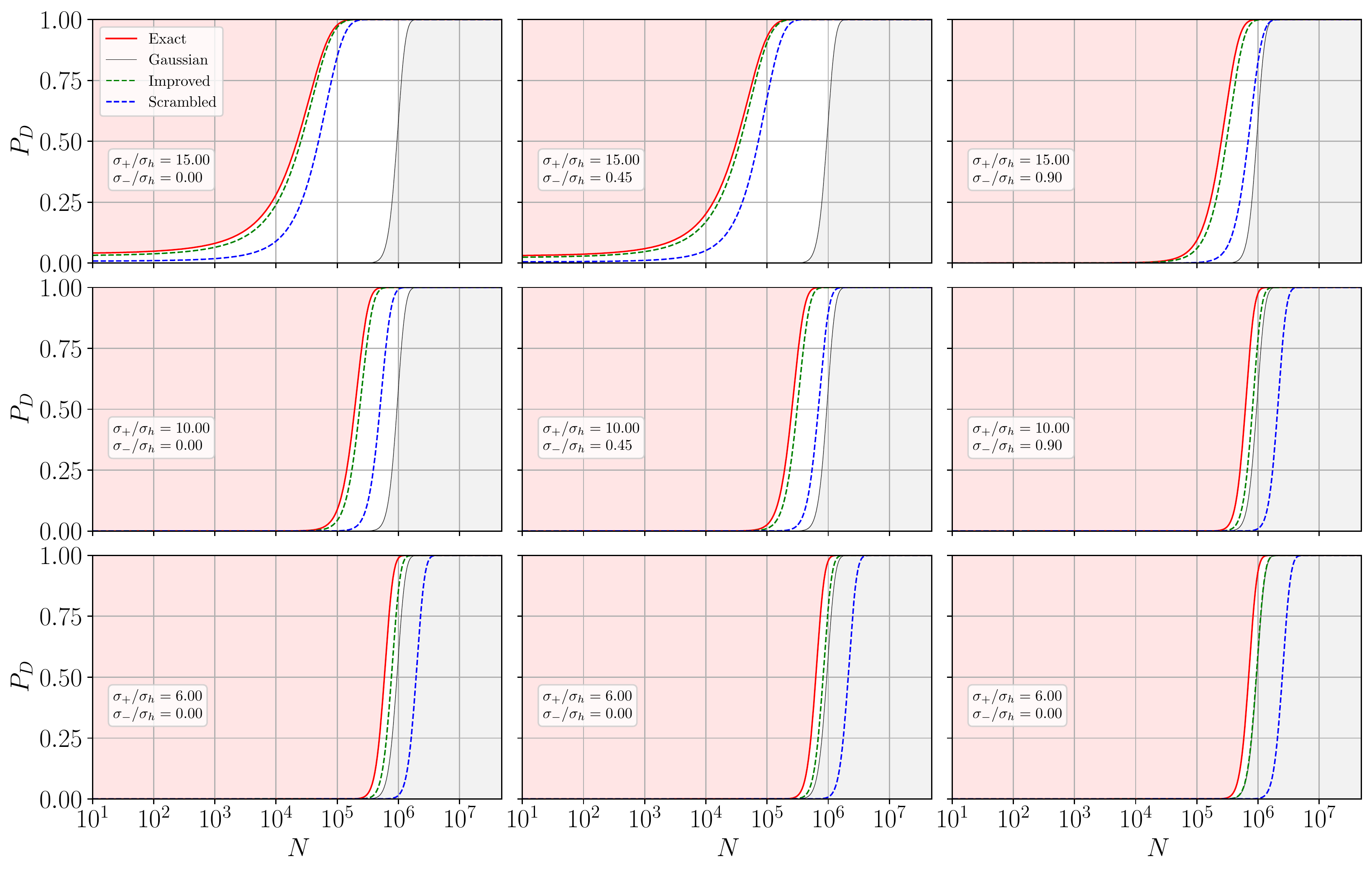}
		\caption{Comparison between the detection probability of the considered DSs as a function of the number of used data $N$. Several diffent signal models are considered, parameterized by $\sigma_+/\sigma_h$ and $\sigma_-/\sigma_h$.
			Here $N_D=4$ and $P_{FA}=10^{-15}$}.
		\label{fig:nd4}
	\end{figure*}

\begin{figure*}
		\centering
		\includegraphics[width=1.84\columnwidth]{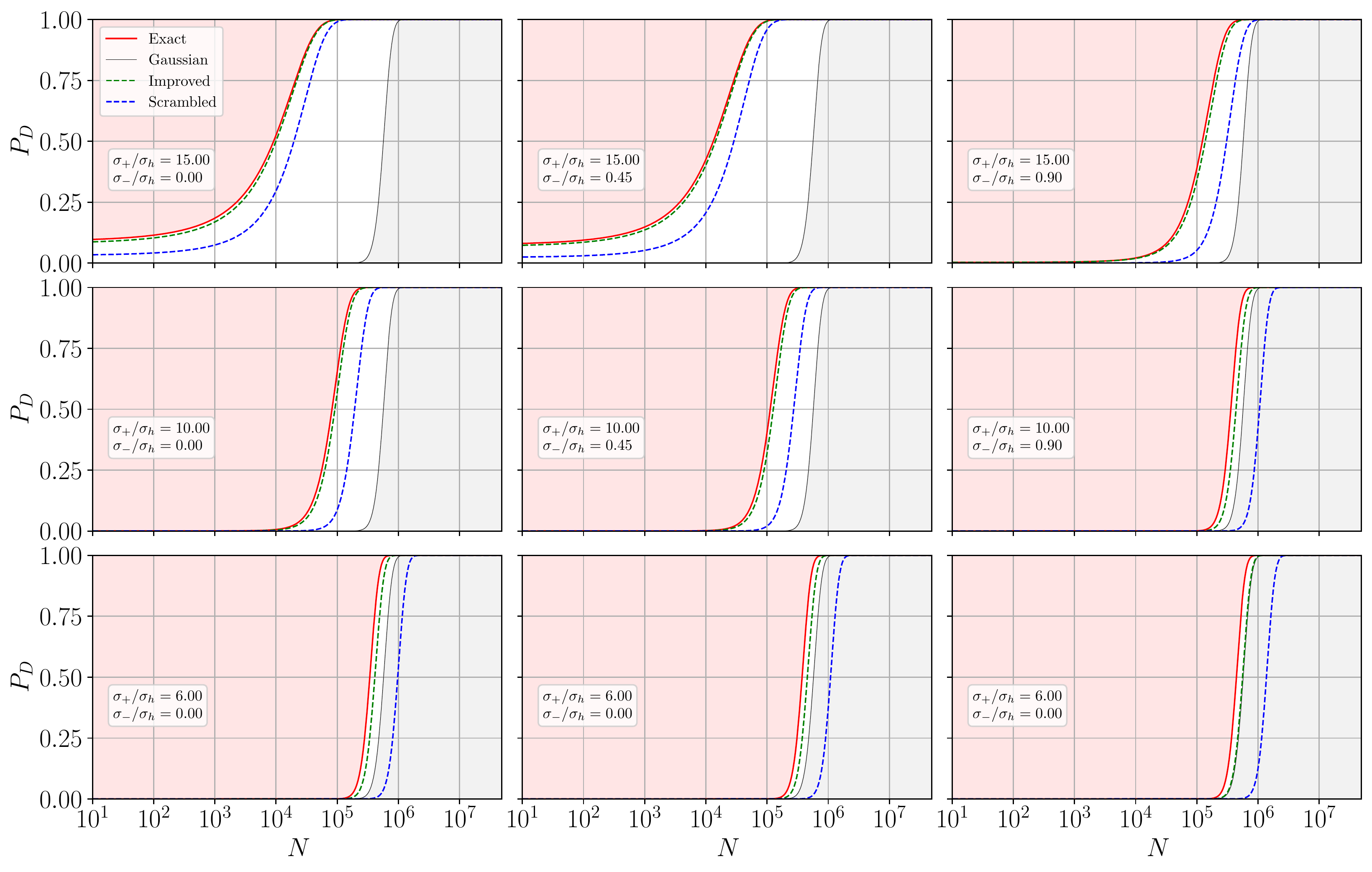}
		\caption{Comparison between the detection probability of the considered DSs as a function of the number of used data $N$. Several diffent signal models are considered, parameterized by $\sigma_+/\sigma_h$ and $\sigma_-/\sigma_h$.
			Here $N_D=5$ and $P_{FA}=10^{-15}$}.
		\label{fig:nd5}
	\end{figure*}

	Performances are presented in Fig.~\ref{fig:nd2},~\ref{fig:nd3},~\ref{fig:nd4} and~\ref{fig:nd5} using Eq.~\eqref{eq:scaledpower} for $N_D =2,3,4$ and $5$, respectively. 
	We plot the detection probability against the number of datapoints N, for $P_{FA}=10^{-15}$. 
	The advantage of the procedure is that it gives precise estimates of the relevant quantities also for very small values of $P_{FA}$, otherwise inaccessible through simulated data.
	On the other hand, the functional form of Eq.~\eqref{eq:scaledpower} is based on the assumption that the statistic is Gaussian, so it should considered reliable only for high enough number of data.
	
	Every figure shows several plots with different choices of $\sigma_+/\sigma_h$ and $\sigma_-/\sigma_h$, to explore different levels of non-Gaussianity, from negligible (bottom right panel) to strong (top left panel). 
	The ``Improved'' DS systematically outperforms the ``Scrambled'' one studied in~\cite{PhysRevD.107.063027}, with greater probabilities of detections for every $N$ and level of signal Gaussianity. 
	In particular, the ``Improved'' performances are always equal to or better than the ``Gaussian'' ones. 
	The ``Improved'' DS does not suffer of the loss of performances affecting the ``Scrambled'' one, even for signals very close to Gaussian.
	
	Remarkably, increasing the number of DSs (i.e. in Fig.~\ref{fig:nd3}) we see that the ``Improved'' DS has performances similar to the ``Exact'' one, showing an important gain in having greater $N_D$.
	The improved DS performances converge to those of the ``Exact'' one for large $N_D$. 
	
	A semi-quantitative understanding of this behavior is obtained as follows: 
	we notice that for a given order $O(u^{2n})$, 
	there are
	\begin{equation}
		N_{n,N_D}=\binom{2n+(N_D-1)}{2n}
	\end{equation}
	possible terms in the Taylor expansion of a DS. 
	The number of terms with even powers of all variables is
	\begin{equation}
		N^{(s)}_{n,N_D}=\binom{n+(N_D-1)}{n}
	\end{equation}
	and the ratio $N^{(s)}_{n,N_D}/N_{n,N_D}$ scales as $N_D^{-n}$ in the large $N_D$ limit:
	\begin{equation}
		\label{eq:ndscaling}
		\frac{N^{(s)}_{n,N_D}}{N_{n,N_D}} 
		=\frac{2^n (2n-1)!!}{\prod_{i=1}^n(N_D+2n-i)}\ .
	\end{equation}
	This suggests that subtracted terms impact less and less the performances when $N_D$ grows larger, the ``Improved'' DS getting closer and closer to the ``Exact'' one.
	We note that a network with $N_D=5$ is not unrealistic \cite{2020LRR....23....3A}, 
	and is already large enough to make the ``Exact'' and ``Improved'' DSs nearly equivalent.  
	
	A further argument in support of such behaviour follows: we rewrite Eq.~\eqref{eq:scaledpower} as
	\begin{align} 
		N_{d} &
		= \!\!
		\left[
		\frac{\sigma_{\Hone}}{\mu_{\Hone}-\mu_{\Hzero}}\bigg\vert_{{N_{d}=1}} \times 
		\right.
		\nonumber \\
		&
		\times \!\!
		\left.
		\left( 
		\frac{\sigma_{\Hzero}}{\sigma_{\Hone}}\bigg\vert_{{N_{d}=1}} 
		\!\!\!\!\!\!\!\!
		\mathrm{ercf}^{-1}(2\pfa)  
		- \mathrm{ercf}^{-1}(2\pd) 
		\right)
		\right]^2 \ ,
	\end{align}
	
	\begin{figure*}
		\centering
		\includegraphics[width=2\columnwidth]{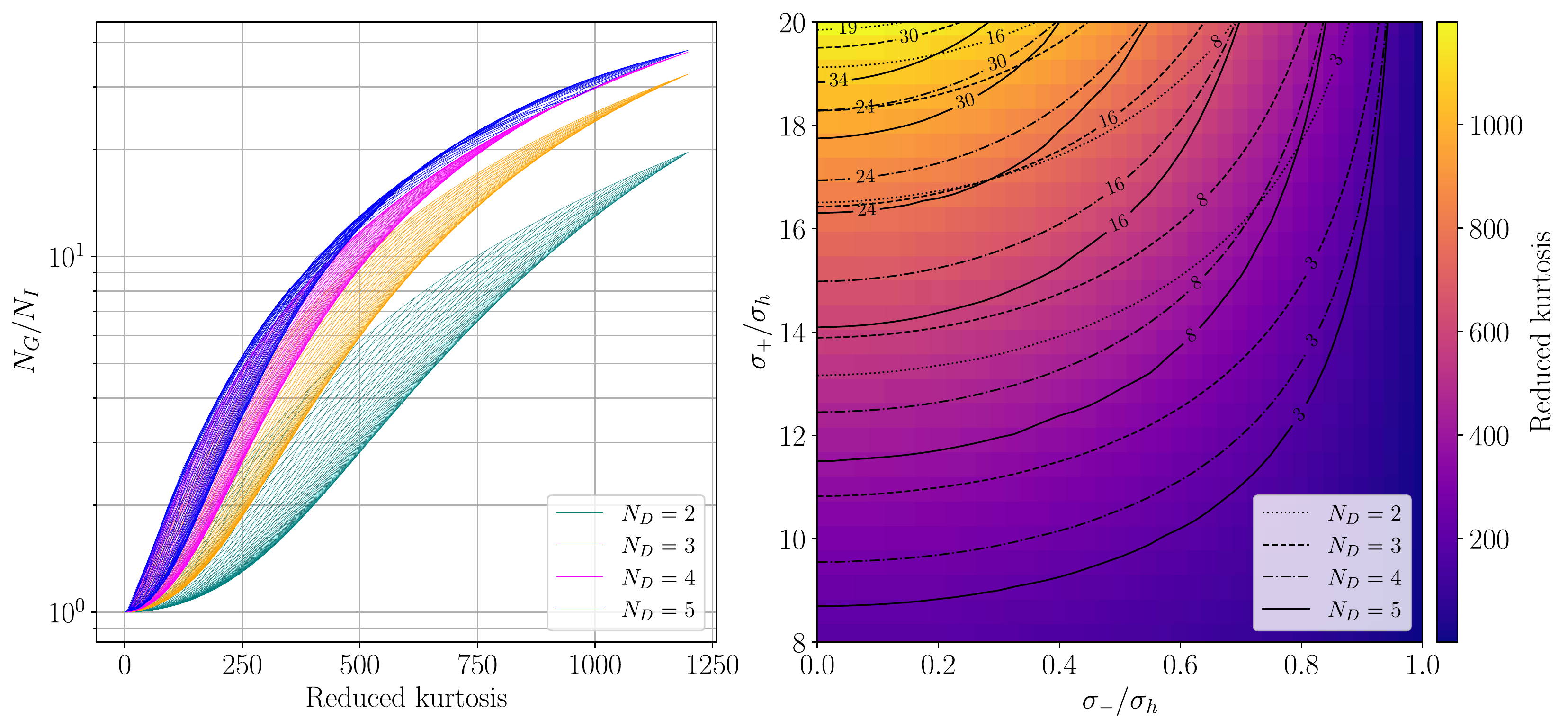}
		\caption{(\textit{Right panel}) Contour levels of the improvement factor $N_{G}/N_{I}$ in a range of the model parameters $\sigma_+/\sigma_h$, $\sigma_-/\sigma_h$.
		Solid, dot-dashed, dashed, and dotted lines denote the improvement factor for $N_D=2,3,4,5$, respectively.
		Pixels shading denotes the model reduced kurtosis. 
		We observe $N_G/N_I$ greater than one, up to a few tens, over the explored parameter space, systematically increasing for increasing number of detectors.
		(\textit{Left panel}) The above performances are shown as a function of the reduced kurtosis and $N_G/N_I$. 
		Uniform grids over the signal parameter space are plotted as teal, orange, purple, and blue meshes for $N_D=2,3,4$ and $5$, respectively.
		We fix as fiducial values $P_D=0.5$ and $P_{FA}=10^{-15}$}.
		\label{fig:density}
	\end{figure*}
	and we define $N_{G}$ and $N_{I}$ its value computed for the ``Gaussian'' and ``Improved'' DSs, respectively. 
	Therefore the ratio $N_{G}/N_{I}$ gives the multiplicative factor for the number of data needed for the ``Gaussian'' DS to reach the same detection probability of the ``Improved'' one, for fixed competing hypotheses and probability of false alarm. 
	
	We show a plot of such ratio at fixed $\pd=0.5$ and $\pfa=10^{-15}$ in Fig.~\ref{fig:density} (right panel) for a set of $N_D$ values. 
	The ratios $\sigma_+/\sigma_h$ and $\sigma_-/\sigma_h$ are shown on the vertical and horizontal axis, respectively. 
	The improved statistic has always better performances than the Gaussian statistic, as $N_{G}/N_{I}$ is globally greater than one. 
	Such improvement grows larger in the upper left corner, where the signal kurtosis is larger, as expected.
	In the right panel of Figure \ref{fig:density} we represent the relation between the reduced kurtosis and $N_{G}/N_{I}$, for a uniform sampling in the parameters $\sigma_+/\sigma_h$, $\sigma_-/\sigma_h$ of the region shown in the right panel. 
	Results for $N_D=2,3,4,5$ can be compared, and we see again that the performance of the ``Improved'' DS relative to the ``Gaussian'' one increase with $N_D$.  
	\gcadd{This is equivalent to a comparison of the time required to achieve a detection at the same level of significance using the two detection statistics.
	From Figure~\ref{fig:density} we see that a confident detection would require a data taking time three times larger with the standard "Gaussian" DS compared with the "Improved" DS we introduced with two detector and a mild non--Gaussianity. Larger improvements are obtained with larger non--Gaussianities and more that two detectors.}
	
	We note that having $N_{G}/N_{I} \geq 1$ is \textit{a priori} expected. 
	In fact, the ``Improved'' version of the optimal statistic is obtained subtracting non-zero mean terms from the ``Exact'' one, which is optimal in the Neyman--Pearson sense. 
	In the Gaussian limit the ``Improved'' statistic becomes exactly the ``Gaussian'' one, as discussed in the previous section (see Eq.~\eqref{eq:gauss3} for a specific example), and can be considered its natural generalization. 
	However, we do not provide here a rigorous proof.
	
	\subsection{Subtraction for realistic backgrounds}
	\label{subsec:realistic}
	
	Given the effectiveness of the new procedure, we explore its application to a realistic non-Gaussian stochastic background. 
	We consider the model
	\begin{equation}
		s_i^{\cal A} = n_i^{\cal A} + h_i^{\cal A} \ ,
	\end{equation}
	where the GW signals measured by each detector are now different, and the joint probability distribution reads
	\begin{equation}
		dP = {\cal D}(h) \prod_{{\cal A},i}  dh_i^{\cal A}
	\end{equation}
	\gcadd{where ${\cal D}(h)$ is some generic function of the data $h_i^{\cal A}$ (a probability density) and 
		$\prod_{{\cal A},i}  dh_i^{\cal A}$ a measure in the space (of dimension $N_D\times N$) of possible signal time series.} 
		
	We first observe that the statistical properties of $h$ are irrelevant, because we only want to obtain the cancellation of \emph{diagonal} terms under the ${\cal H}_0$ hypothesis. 
	However, realistic detector noises can have non trivial time-domain correlations, and non-Gaussian contaminations are expected.
	A generic term of a Taylor expansion of the DSs with degree $\kappa_1+\cdots+\kappa_{N_D}$ reads
	\begin{equation}
		\underbrace{\left( n^{1}_{i_{1,1}} \cdots    n^{1}_{i_{1,\kappa_1}} \right)}_{\text{Detector 1}} \cdots \underbrace{\left( n^{N_D}_{i_{N_D,1}} \cdots    n^{N_D}_{i_{N_D,\kappa_{N_D}}} \right)}_{\text{Detector $N_D$}} \ ,
	\end{equation}
	and its expectation is
	\begin{equation}
		\left< n^{1}_{i_{1,1}} \cdots    n^{1}_{i_{1,\kappa_1}} \right> \cdots \left< n^{N_D}_{i_{N_D,1}} \cdots    n^{N_D}_{i_{N_D,\kappa_{N_D}}} \right>
	\end{equation}
	as we assume again statistically independent noises across different detectors.
	After applying the subtraction procedure defined in~Eq.~\eqref{eq:subtraction} we get
	\begin{multline}
		\left( 
		n^{1}_{i_{1,1}} 
		\cdots    
		n^{1}_{i_{1,\kappa_1}} 
		\right) 
		\cdots 
		\left( 
		n^{N_D}_{i_{N_D,1}} 
		\cdots    
		n^{N_D}_{i_{N_D,\kappa_{N_D}}} 
		\right) 
		\rightarrow \nonumber \\
		\left(
		1-\frac{1}{2^{N_D}} 
		\sum_{\underline{\varepsilon}} 
		\varepsilon_{1}^{\kappa_1} 
		\cdots 
		\varepsilon_{N_D}^{\kappa_{N_D}} 
		\right)  \nonumber \\
		\times \left( n^{1}_{i_{1,1}} 
		\cdots    
		n^{1}_{i_{1,\kappa_1}} 
		\right) 
		\cdots 
		\left( n^{N_D}_{i_{N_D,1}} 
		\cdots    n^{N_D}_{i_{N_D,\kappa_{N_D}}} 
		\right) \ .
	\end{multline}
	As in the simplified case, a given term is cancelled if all the $\kappa_{\cal A}$s are even and is preserved otherwise. 
	The expectation values of all the preserved contributions are zero, unless some odd-order noise momenta are non null. 
	Therefore, the subtraction procedure is effective in a very general sense, and is applicable also when non-Gaussian contaminations (without non-zero odd momenta) are present. 
	\gcrep{In particular this procedure can be incorporated easily in the general DS defined in Eq.~(17) of~\cite{PhysRevD.107.063027}.
	However, it should be noted that non-Gaussian noise contributions can change the expectation value of the statistic under ${\cal H}_1$.}{
	However, it should be noted that non-Gaussian noise contributions can change the expectation value of the statistic under ${\cal H}_1$. In particular this procedure can be incorporated easily in the general DS defined in Eq.~(17) of~\cite{PhysRevD.107.063027}.}
	
	
	\section{Conclusions and perspectives}
	\label{sec:Conclusions-and-perspectives}
	
	The ``Improved'' DS introduced in this paper can be seen as the natural generalization to the non--Gaussian case of the optimal statistic used for a Gaussian stochastic background.
	\gcadd{As a study of a realistic model is computationally expensive, we performed an exploration based on a simplified toy model, and focused in this manuscript on our findings.} 
	The toy model we used to quantify its performances is simplified. \gcadd{In particular, it completely neglects the non trivial time structure of the signals associated to a realistic event.} However we showed that the core procedure to remove \emph{diagonal} contributions is applicable to the general case, since it is insensitive to time or frequency structure, and only performs a transformation in the detector space.
	
	Performance results are promising\gcrep{, and}{:} we expect that the improvement with respect to the ``Gaussian'' DS will be preserved in a realistic situation. 
	We observed that the ``Improved'' DS becomes nearly equivalent to the ``Exact'' one for a large enough (although reasonable) number of detectors in the network.
	\gcadd{For completeness, it should be noted that the improvement that can be obtained in a real scenario depends on the details of the background non--Gaussianity, so one should extrapolate with care our findings, and ultimately support them through direct investigation.}
	
	\gcadd{From a practical point of view, } the main difference in a realistic scenario \gcadd{where no analytical expression for the Improved DS is available,} arises from the computational cost in evaluating it as discussed in \cite{PhysRevD.107.063027} (e.g. the discussion following Eq.~(17) therein), which requires an efficient numerical procedure. 
	The computational cost is related to the number of configurations needed for an accurate statistic estimate, which we investigated in previous work using importance sampling procedure. 
	On this ground, we expect a large \gcrep{number of detectors}{ $N_D$} to help in reducing the effective computational cost.
	
	Computational cost estimate of the method in a realistic case is a key step that needs to be fully addressed to \gcrep{assess}{understand} the applicability of our method. 
	We argue than an expensive, although manageable, computational cost would certainly be justified by the prospect of an earlier and more confident detection of an astrophysical stochastic background, e.g. generated by a superposition of binary coalescence events, within reach in the near future~\cite{2021PhRvD.104b2004A}. \gcadd{We expect that no major additional obstructions will forbid the application of the method to other detectors, such as LISA or pulsar timing arrays. }

	\begin{acknowledgments}
		RB acknowledges support through the Italian Space Agency grant \emph{Phase A activity for LISA mission, Agreement n. 2017-29-H.0, CUP F62F17000290005}.
		
		\textit{Software}:
		We acknowledge usage of 
		\textsc{Mathematica}~\cite{Mathematica}
		and of the following
		\textsc{Python}~\cite{10.5555/1593511}
		packages for the analysis, post-processing and production of results throughout:
		%\textsc{CPNest}~\cite{2021zndo...4470001V},
		\textsc{matplotlib}~\cite{2007CSE.....9...90H},
		\textsc{numpy}~\cite{2020Natur.585..357H},
		\textsc{scipy}~\cite{2020NatMe..17..261V}.
	\end{acknowledgments}
	
	\vfill
	\bibliographystyle{apsrev4-2}
	\bibliography{biblio}

\begin{thebibliography}{22}%
\makeatletter
\providecommand \@ifxundefined [1]{%
 \@ifx{#1\undefined}
}%
\providecommand \@ifnum [1]{%
 \ifnum #1\expandafter \@firstoftwo
 \else \expandafter \@secondoftwo
 \fi
}%
\providecommand \@ifx [1]{%
 \ifx #1\expandafter \@firstoftwo
 \else \expandafter \@secondoftwo
 \fi
}%
\providecommand \natexlab [1]{#1}%
\providecommand \enquote  [1]{``#1''}%
\providecommand \bibnamefont  [1]{#1}%
\providecommand \bibfnamefont [1]{#1}%
\providecommand \citenamefont [1]{#1}%
\providecommand \href@noop [0]{\@secondoftwo}%
\providecommand \href [0]{\begingroup \@sanitize@url \@href}%
\providecommand \@href[1]{\@@startlink{#1}\@@href}%
\providecommand \@@href[1]{\endgroup#1\@@endlink}%
\providecommand \@sanitize@url [0]{\catcode `\\12\catcode `\$12\catcode
  `\&12\catcode `\#12\catcode `\^12\catcode `\_12\catcode `\%12\relax}%
\providecommand \@@startlink[1]{}%
\providecommand \@@endlink[0]{}%
\providecommand \url  [0]{\begingroup\@sanitize@url \@url }%
\providecommand \@url [1]{\endgroup\@href {#1}{\urlprefix }}%
\providecommand \urlprefix  [0]{URL }%
\providecommand \Eprint [0]{\href }%
\providecommand \doibase [0]{https://doi.org/}%
\providecommand \selectlanguage [0]{\@gobble}%
\providecommand \bibinfo  [0]{\@secondoftwo}%
\providecommand \bibfield  [0]{\@secondoftwo}%
\providecommand \translation [1]{[#1]}%
\providecommand \BibitemOpen [0]{}%
\providecommand \bibitemStop [0]{}%
\providecommand \bibitemNoStop [0]{.\EOS\space}%
\providecommand \EOS [0]{\spacefactor3000\relax}%
\providecommand \BibitemShut  [1]{\csname bibitem#1\endcsname}%
\let\auto@bib@innerbib\@empty
%</preamble>
\bibitem [{\citenamefont {{Abbott}}\ \emph {et~al.}(2021)\citenamefont
  {{Abbott}}, \citenamefont {{Abbott}}, \citenamefont {{Abraham}},
  \citenamefont {{Acernese}} \emph {et~al.}}]{2021PhRvD.104b2004A}%
  \BibitemOpen
  \bibfield  {author} {\bibinfo {author} {\bibfnamefont {R.}~\bibnamefont
  {{Abbott}}}, \bibinfo {author} {\bibfnamefont {T.~D.}\ \bibnamefont
  {{Abbott}}}, \bibinfo {author} {\bibfnamefont {S.}~\bibnamefont {{Abraham}}},
  \bibinfo {author} {\bibfnamefont {F.}~\bibnamefont {{Acernese}}}, \emph
  {et~al.},\ }\href {https://doi.org/10.1103/PhysRevD.104.022004} {\bibfield
  {journal} {\bibinfo  {journal} {\prd}\ }\textbf {\bibinfo {volume} {104}},\
  \bibinfo {eid} {022004} (\bibinfo {year} {2021})},\ \Eprint
  {https://arxiv.org/abs/2101.12130} {arXiv:2101.12130 [gr-qc]} \BibitemShut
  {NoStop}%
\bibitem [{\citenamefont {Regimbau}(2011)}]{TaniaRegimbau_2011}%
  \BibitemOpen
  \bibfield  {author} {\bibinfo {author} {\bibfnamefont {T.}~\bibnamefont
  {Regimbau}},\ }\href {https://doi.org/10.1088/1674-4527/11/4/001} {\bibfield
  {journal} {\bibinfo  {journal} {Research in Astronomy and Astrophysics}\
  }\textbf {\bibinfo {volume} {11}},\ \bibinfo {pages} {369} (\bibinfo {year}
  {2011})}\BibitemShut {NoStop}%
\bibitem [{\citenamefont {Damour}\ and\ \citenamefont
  {Vilenkin}(2000)}]{PhysRevLett.85.3761}%
  \BibitemOpen
  \bibfield  {author} {\bibinfo {author} {\bibfnamefont {T.}~\bibnamefont
  {Damour}}\ and\ \bibinfo {author} {\bibfnamefont {A.}~\bibnamefont
  {Vilenkin}},\ }\href {https://doi.org/10.1103/PhysRevLett.85.3761} {\bibfield
   {journal} {\bibinfo  {journal} {Phys. Rev. Lett.}\ }\textbf {\bibinfo
  {volume} {85}},\ \bibinfo {pages} {3761} (\bibinfo {year}
  {2000})}\BibitemShut {NoStop}%
\bibitem [{\citenamefont {{The LIGO Scientific Collaboration}}\ \emph
  {et~al.}(2021)\citenamefont {{The LIGO Scientific Collaboration}},
  \citenamefont {{the Virgo Collaboration}},\ and\ \citenamefont {{the KAGRA
  Collaboration}}}]{2021arXiv211103634T}%
  \BibitemOpen
  \bibfield  {author} {\bibinfo {author} {\bibnamefont {{The LIGO Scientific
  Collaboration}}}, \bibinfo {author} {\bibnamefont {{the Virgo
  Collaboration}}},\ and\ \bibinfo {author} {\bibnamefont {{the KAGRA
  Collaboration}}},\ }\href {https://doi.org/10.48550/arXiv.2111.03634}
  {\bibfield  {journal} {\bibinfo  {journal} {arXiv e-prints}\ ,\ \bibinfo
  {eid} {arXiv:2111.03634}} (\bibinfo {year} {2021})},\ \Eprint
  {https://arxiv.org/abs/2111.03634} {arXiv:2111.03634 [astro-ph.HE]}
  \BibitemShut {NoStop}%
\bibitem [{\citenamefont {Drasco}\ and\ \citenamefont
  {Flanagan}(2003)}]{Drasco:2002yd}%
  \BibitemOpen
  \bibfield  {author} {\bibinfo {author} {\bibfnamefont {S.}~\bibnamefont
  {Drasco}}\ and\ \bibinfo {author} {\bibfnamefont {E.~E.}\ \bibnamefont
  {Flanagan}},\ }\href {https://doi.org/10.1103/PhysRevD.67.082003} {\bibfield
  {journal} {\bibinfo  {journal} {Phys. Rev.}\ }\textbf {\bibinfo {volume}
  {D67}},\ \bibinfo {pages} {082003} (\bibinfo {year} {2003})},\ \Eprint
  {https://arxiv.org/abs/gr-qc/0210032} {arXiv:gr-qc/0210032 [gr-qc]}
  \BibitemShut {NoStop}%
%%CITATION = GR-QC/0210032;%%
\bibitem [{\citenamefont {{Thrane}}(2013)}]{2013PhRvD..87d3009T}%
  \BibitemOpen
  \bibfield  {author} {\bibinfo {author} {\bibfnamefont {E.}~\bibnamefont
  {{Thrane}}},\ }\href {https://doi.org/10.1103/PhysRevD.87.043009} {\bibfield
  {journal} {\bibinfo  {journal} {\prd}\ }\textbf {\bibinfo {volume} {87}},\
  \bibinfo {eid} {043009} (\bibinfo {year} {2013})},\ \Eprint
  {https://arxiv.org/abs/1301.0263} {arXiv:1301.0263 [astro-ph.IM]}
  \BibitemShut {NoStop}%
\bibitem [{\citenamefont {{Smith}}\ and\ \citenamefont
  {{Thrane}}(2018)}]{Smith:2018PhRvX...8b1019S}%
  \BibitemOpen
  \bibfield  {author} {\bibinfo {author} {\bibfnamefont {R.}~\bibnamefont
  {{Smith}}}\ and\ \bibinfo {author} {\bibfnamefont {E.}~\bibnamefont
  {{Thrane}}},\ }\href {https://doi.org/10.1103/PhysRevX.8.021019} {\bibfield
  {journal} {\bibinfo  {journal} {Physical Review X}\ }\textbf {\bibinfo
  {volume} {8}},\ \bibinfo {eid} {021019} (\bibinfo {year} {2018})},\ \Eprint
  {https://arxiv.org/abs/1712.00688} {arXiv:1712.00688 [gr-qc]} \BibitemShut
  {NoStop}%
\bibitem [{\citenamefont {{Martellini}}\ and\ \citenamefont
  {{Regimbau}}(2014)}]{2014PhRvD..89l4009M}%
  \BibitemOpen
  \bibfield  {author} {\bibinfo {author} {\bibfnamefont {L.}~\bibnamefont
  {{Martellini}}}\ and\ \bibinfo {author} {\bibfnamefont {T.}~\bibnamefont
  {{Regimbau}}},\ }\href {https://doi.org/10.1103/PhysRevD.89.124009}
  {\bibfield  {journal} {\bibinfo  {journal} {\prd}\ }\textbf {\bibinfo
  {volume} {89}},\ \bibinfo {eid} {124009} (\bibinfo {year} {2014})},\ \Eprint
  {https://arxiv.org/abs/1405.5775} {arXiv:1405.5775 [astro-ph.CO]}
  \BibitemShut {NoStop}%
\bibitem [{\citenamefont {{Seto}}(2008)}]{2008ApJ...683L..95S}%
  \BibitemOpen
  \bibfield  {author} {\bibinfo {author} {\bibfnamefont {N.}~\bibnamefont
  {{Seto}}},\ }\href {https://doi.org/10.1086/591847} {\bibfield  {journal}
  {\bibinfo  {journal} {\apjl}\ }\textbf {\bibinfo {volume} {683}},\ \bibinfo
  {pages} {L95} (\bibinfo {year} {2008})},\ \Eprint
  {https://arxiv.org/abs/0807.1151} {arXiv:0807.1151 [astro-ph]} \BibitemShut
  {NoStop}%
\bibitem [{\citenamefont {{Seto}}(2009)}]{2009PhRvD..80d3003S}%
  \BibitemOpen
  \bibfield  {author} {\bibinfo {author} {\bibfnamefont {N.}~\bibnamefont
  {{Seto}}},\ }\href {https://doi.org/10.1103/PhysRevD.80.043003} {\bibfield
  {journal} {\bibinfo  {journal} {\prd}\ }\textbf {\bibinfo {volume} {80}},\
  \bibinfo {eid} {043003} (\bibinfo {year} {2009})},\ \Eprint
  {https://arxiv.org/abs/0908.0228} {arXiv:0908.0228 [gr-qc]} \BibitemShut
  {NoStop}%
\bibitem [{\citenamefont {Romano}\ and\ \citenamefont
  {Cornish}(2017)}]{Romano:2016dpx}%
  \BibitemOpen
  \bibfield  {author} {\bibinfo {author} {\bibfnamefont {J.~D.}\ \bibnamefont
  {Romano}}\ and\ \bibinfo {author} {\bibfnamefont {N.~J.}\ \bibnamefont
  {Cornish}},\ }\href {https://doi.org/10.1007/s41114-017-0004-1} {\bibfield
  {journal} {\bibinfo  {journal} {Living Rev. Rel.}\ }\textbf {\bibinfo
  {volume} {20}},\ \bibinfo {pages} {2} (\bibinfo {year} {2017})},\ \Eprint
  {https://arxiv.org/abs/1608.06889} {arXiv:1608.06889 [gr-qc]} \BibitemShut
  {NoStop}%
%%CITATION = ARXIV:1608.06889;%%
\bibitem [{\citenamefont {Buscicchio}\ \emph {et~al.}(2023)\citenamefont
  {Buscicchio}, \citenamefont {Ain}, \citenamefont {Ballelli}, \citenamefont
  {Cella},\ and\ \citenamefont {Patricelli}}]{PhysRevD.107.063027}%
  \BibitemOpen
  \bibfield  {author} {\bibinfo {author} {\bibfnamefont {R.}~\bibnamefont
  {Buscicchio}}, \bibinfo {author} {\bibfnamefont {A.}~\bibnamefont {Ain}},
  \bibinfo {author} {\bibfnamefont {M.}~\bibnamefont {Ballelli}}, \bibinfo
  {author} {\bibfnamefont {G.}~\bibnamefont {Cella}},\ and\ \bibinfo {author}
  {\bibfnamefont {B.}~\bibnamefont {Patricelli}},\ }\href
  {https://doi.org/10.1103/PhysRevD.107.063027} {\bibfield  {journal} {\bibinfo
   {journal} {Phys. Rev. D}\ }\textbf {\bibinfo {volume} {107}},\ \bibinfo
  {pages} {063027} (\bibinfo {year} {2023})}\BibitemShut {NoStop}%
\bibitem [{\citenamefont {{Allen}}\ and\ \citenamefont
  {{Romano}}(1999)}]{1999PhRvD..59j2001A}%
  \BibitemOpen
  \bibfield  {author} {\bibinfo {author} {\bibfnamefont {B.}~\bibnamefont
  {{Allen}}}\ and\ \bibinfo {author} {\bibfnamefont {J.~D.}\ \bibnamefont
  {{Romano}}},\ }\href {https://doi.org/10.1103/PhysRevD.59.102001} {\bibfield
  {journal} {\bibinfo  {journal} {\prd}\ }\textbf {\bibinfo {volume} {59}},\
  \bibinfo {eid} {102001} (\bibinfo {year} {1999})},\ \Eprint
  {https://arxiv.org/abs/gr-qc/9710117} {arXiv:gr-qc/9710117 [gr-qc]}
  \BibitemShut {NoStop}%
\bibitem [{\citenamefont {{Renzini}}\ \emph {et~al.}(2022)\citenamefont
  {{Renzini}}, \citenamefont {{Goncharov}}, \citenamefont {{Jenkins}},\ and\
  \citenamefont {{Meyers}}}]{2022Galax..10...34R}%
  \BibitemOpen
  \bibfield  {author} {\bibinfo {author} {\bibfnamefont {A.~I.}\ \bibnamefont
  {{Renzini}}}, \bibinfo {author} {\bibfnamefont {B.}~\bibnamefont
  {{Goncharov}}}, \bibinfo {author} {\bibfnamefont {A.~C.}\ \bibnamefont
  {{Jenkins}}},\ and\ \bibinfo {author} {\bibfnamefont {P.~M.}\ \bibnamefont
  {{Meyers}}},\ }\href {https://doi.org/10.3390/galaxies10010034} {\bibfield
  {journal} {\bibinfo  {journal} {Galaxies}\ }\textbf {\bibinfo {volume}
  {10}},\ \bibinfo {pages} {34} (\bibinfo {year} {2022})},\ \Eprint
  {https://arxiv.org/abs/2202.00178} {arXiv:2202.00178 [gr-qc]} \BibitemShut
  {NoStop}%
\bibitem [{\citenamefont {{Neyman}}\ and\ \citenamefont
  {{Pearson}}(1933)}]{1933RSPTA.231..289N}%
  \BibitemOpen
  \bibfield  {author} {\bibinfo {author} {\bibfnamefont {J.}~\bibnamefont
  {{Neyman}}}\ and\ \bibinfo {author} {\bibfnamefont {E.~S.}\ \bibnamefont
  {{Pearson}}},\ }\href {https://doi.org/10.1098/rsta.1933.0009} {\bibfield
  {journal} {\bibinfo  {journal} {Philosophical Transactions of the Royal
  Society of London Series A}\ }\textbf {\bibinfo {volume} {231}},\ \bibinfo
  {pages} {289} (\bibinfo {year} {1933})}\BibitemShut {NoStop}%
\bibitem [{\citenamefont {van Kampen}(1992)}]{Kampen1992}%
  \BibitemOpen
  \bibfield  {author} {\bibinfo {author} {\bibfnamefont {N.}~\bibnamefont {van
  Kampen}},\ }\href@noop {} {\emph {\bibinfo {title} {{S}tochastic {P}rocesses
  in {P}hysics and {C}hemistry}}}\ (\bibinfo  {publisher} {Elsevier Science
  Publishers, Amsterdam},\ \bibinfo {year} {1992})\BibitemShut {NoStop}%
\bibitem [{\citenamefont {{Abbott}}\ \emph {et~al.}(2020)\citenamefont
  {{Abbott}} \emph {et~al.}}]{2020LRR....23....3A}%
  \BibitemOpen
  \bibfield  {author} {\bibinfo {author} {\bibfnamefont {B.~P.}\ \bibnamefont
  {{Abbott}}} \emph {et~al.},\ }\href
  {https://doi.org/10.1007/s41114-020-00026-9} {\bibfield  {journal} {\bibinfo
  {journal} {Living Reviews in Relativity}\ }\textbf {\bibinfo {volume} {23}},\
  \bibinfo {eid} {3} (\bibinfo {year} {2020})}\BibitemShut {NoStop}%
\bibitem [{\citenamefont {Inc.}(2022)}]{Mathematica}%
  \BibitemOpen
  \bibfield  {author} {\bibinfo {author} {\bibfnamefont {W.~R.}\ \bibnamefont
  {Inc.}},\ }\href {https://www.wolfram.com/mathematica} {\bibinfo {title}
  {Mathematica, {V}ersion 13.1}} (\bibinfo {year} {2022}),\ \bibinfo {note}
  {{C}hampaign, IL}\BibitemShut {NoStop}%
\bibitem [{\citenamefont {Van~Rossum}\ and\ \citenamefont
  {Drake}(2009)}]{10.5555/1593511}%
  \BibitemOpen
  \bibfield  {author} {\bibinfo {author} {\bibfnamefont {G.}~\bibnamefont
  {Van~Rossum}}\ and\ \bibinfo {author} {\bibfnamefont {F.~L.}\ \bibnamefont
  {Drake}},\ }\href@noop {} {\emph {\bibinfo {title} {Python 3 Reference
  Manual}}}\ (\bibinfo  {publisher} {CreateSpace},\ \bibinfo {address} {Scotts
  Valley, CA},\ \bibinfo {year} {2009})\BibitemShut {NoStop}%
\bibitem [{\citenamefont {{Hunter}}(2007)}]{2007CSE.....9...90H}%
  \BibitemOpen
  \bibfield  {author} {\bibinfo {author} {\bibfnamefont {J.~D.}\ \bibnamefont
  {{Hunter}}},\ }\href {https://doi.org/10.1109/MCSE.2007.55} {\bibfield
  {journal} {\bibinfo  {journal} {Computing in Science and Engineering}\
  }\textbf {\bibinfo {volume} {9}},\ \bibinfo {pages} {90} (\bibinfo {year}
  {2007})}\BibitemShut {NoStop}%
\bibitem [{\citenamefont {{Harris}}\ \emph {et~al.}(2020)\citenamefont
  {{Harris}}, \citenamefont {{Millman}}, \citenamefont {{van der Walt}},
  \citenamefont {{Gommers}} \emph {et~al.}}]{2020Natur.585..357H}%
  \BibitemOpen
  \bibfield  {author} {\bibinfo {author} {\bibfnamefont {C.~R.}\ \bibnamefont
  {{Harris}}}, \bibinfo {author} {\bibfnamefont {K.~J.}\ \bibnamefont
  {{Millman}}}, \bibinfo {author} {\bibfnamefont {S.~J.}\ \bibnamefont {{van
  der Walt}}}, \bibinfo {author} {\bibfnamefont {R.}~\bibnamefont {{Gommers}}},
  \emph {et~al.},\ }\href {https://doi.org/10.1038/s41586-020-2649-2}
  {\bibfield  {journal} {\bibinfo  {journal} {\nat}\ }\textbf {\bibinfo
  {volume} {585}},\ \bibinfo {pages} {357} (\bibinfo {year} {2020})},\ \Eprint
  {https://arxiv.org/abs/2006.10256} {arXiv:2006.10256 [cs.MS]} \BibitemShut
  {NoStop}%
\bibitem [{\citenamefont {{Virtanen}}\ \emph {et~al.}(2020)\citenamefont
  {{Virtanen}}, \citenamefont {{Gommers}}, \citenamefont {{Oliphant}},
  \citenamefont {{Haberland}} \emph {et~al.}}]{2020NatMe..17..261V}%
  \BibitemOpen
  \bibfield  {author} {\bibinfo {author} {\bibfnamefont {P.}~\bibnamefont
  {{Virtanen}}}, \bibinfo {author} {\bibfnamefont {R.}~\bibnamefont
  {{Gommers}}}, \bibinfo {author} {\bibfnamefont {T.~E.}\ \bibnamefont
  {{Oliphant}}}, \bibinfo {author} {\bibfnamefont {M.}~\bibnamefont
  {{Haberland}}}, \emph {et~al.},\ }\href
  {https://doi.org/10.1038/s41592-019-0686-2} {\bibfield  {journal} {\bibinfo
  {journal} {Nature Methods}\ }\textbf {\bibinfo {volume} {17}},\ \bibinfo
  {pages} {261} (\bibinfo {year} {2020})},\ \Eprint
  {https://arxiv.org/abs/1907.10121} {arXiv:1907.10121 [cs.MS]} \BibitemShut
  {NoStop}%
\end{thebibliography}%
\end{document}